 \documentclass[aps,pre,twocolumn,superscriptaddress,floatfix]{revtex4}
\usepackage{graphicx,bm,amssymb,amsmath,dcolumn,hyperref}
\usepackage{verbatim}
\usepackage{color}
\usepackage{epstopdf}
\usepackage{subfigure}
\usepackage{mathrsfs}
\usepackage{amssymb}

\begin{document}
\newcommand{\be}{\begin{equation}}
\newcommand{\ee}{\end{equation}}
\newcommand{\half}{\frac{1}{2}}
\newcommand{\ith}{^{(i)}}
\newcommand{\im}{^{(i-1)}}
\newcommand{\gae}
{\,\hbox{\lower0.5ex\hbox{$\sim$}\llap{\raise0.5ex\hbox{$>$}}}\,}
\newcommand{\lae}
{\,\hbox{\lower0.5ex\hbox{$\sim$}\llap{\raise0.5ex\hbox{$<$}}}\,}

\definecolor{blue}{rgb}{0,0,1}
\definecolor{red}{rgb}{1,0,0}
\definecolor{green}{rgb}{0,1,0}
\newcommand{\blue}[1]{\textcolor{blue}{#1}}
\newcommand{\red}[1]{\textcolor{red}{#1}}
\newcommand{\green}[1]{\textcolor{green}{#1}}

\newcommand{\yi}{y_{\rm i}}
\newcommand{\pc}{p_{\rm c}}
\newcommand{\PB}{P_{\rm B}}

\newcommand{\calA}{ {\mathcal A}}
\newcommand{\calC}{ {\mathcal C}}
\newcommand{\calE}{ {\mathcal E}}
\newcommand{\calG}{ {\mathcal G}}
\newcommand{\calH}{ {\mathcal H}}
\newcommand{\calO}{ {\mathcal O}}
\newcommand{\calS}{ {\mathcal S}}
\newcommand{\calZ}{ {\mathcal Z}}

\newcommand{\arho}{ {\overline{\rho}}}
\title{Critical polynomials in the nonplanar and continuum percolation models}

\author{Wenhui Xu}
\affiliation{School of Physics and Materials Science, Anhui University, Hefei, Anhui 230601, China}
\affiliation{Hefei National Laboratory for Physical Sciences at the Microscale and
Department of Modern Physics, University of Science and Technology of China,
Hefei, Anhui 230026, China}

\author{Junfeng Wang}
\email{wangjf@hfut.edu.cn}
\affiliation{School of Electronic Science and Applied Physics, Hefei University
of Technology, Hefei, Anhui 230009, China}

\author{Hao Hu}
\email{huhao@ahu.edu.cn}
\affiliation{School of Physics and Materials Science, Anhui University, Hefei, Anhui 230601, China}

\author{Youjin Deng}
\email{yjdeng@ustc.edu.cn}
\affiliation{Hefei National Laboratory for Physical Sciences at the Microscale and
Department of Modern Physics, University of Science and Technology of China,
Hefei, Anhui 230026, China}
\affiliation{Department of Physics and Electronic Information Engineering, Minjiang University, Fuzhou, Fujian 350108, China}

\date{\today}

\begin{abstract}
	Exact or precise thresholds have been intensively studied since the introduction of the percolation model.
	Recently the critical polynomial $\PB(p,L)$ was introduced for planar-lattice percolation models, where $p$ is the occupation probability and $L$ is the linear system size. The solution of $\PB = 0$ can reproduce all known exact thresholds and leads to unprecedented estimates for thresholds of unsolved planar-lattice models. 
	In two dimensions, assuming the universality of $\PB$, we use it to study a nonplanar lattice model, i.e., the equivalent-neighbor lattice bond percolation, and the continuum percolation of identical penetrable disks, by Monte Carlo simulations and finite-size scaling analysis.
	It is found that, in comparison with other quantities, $\PB$ suffers much less from finite-size corrections. 
	As a result, we obtain a series of high-precision thresholds $p_c(z)$ as a function of coordination number $z$ for equivalent-neighbor percolation with $z$ up to O$(10^5)$, and clearly confirm the asymptotic behavior $zp_c-1 \sim 1/\sqrt{z}$ for $z \rightarrow \infty$.
	For the continuum percolation model, we surprisingly observe that the finite-size correction in $P_{\rm B}$ is unobservable within uncertainty O$(10^{-5})$ as long as $L \geq 3$. The estimated threshold number density of disks is $\rho_c = 1.436\,325\,05(10)$, slightly below the most recent result $\rho_c = 1.436\,325\,45(8)$ 
	of Mertens and Moore obtained by other means. 
	Our work suggests that the critical polynomial method can be a powerful tool for studying nonplanar and continuum systems in statistical mechanics.
\end{abstract}
\pacs{}
\maketitle 

\section{Introduction}
\label{sec1}

Percolation theory~\cite{perc} has been extensively studied for more than $60$ years since it was first proposed by Broadbent and Hammersley~\cite{perc0}. It concerns the formation of connected components in random systems, and is one of the simplest examples of phase transitions. Despite the simplicity of its definition, the calculation of percolation thresholds is a very challenging problem. For the convenience of readers, we shall briefly recall some of the methods for analytically solving percolation thresholds in the past $60$ years.

In the early years, only a few special classes of two-dimensional lattices could be exactly solved by using duality or matching properties of the lattices. 
For a given planar lattice $\mathcal{L}$, the dual lattice $\mathcal{L}^*$ can be obtained by doing the following: (i) On each face of $\mathcal{L}$, place a vertex which serves as a vertex of $\mathcal{L}^*$; (ii) For any two vertices of $\mathcal{L}^*$, add an edge between them if the corresponding two faces of $\mathcal{L}$ have a common edge. For bond percolation, the thresholds of a lattice $\mathcal{L}$ and its dual lattice $\mathcal{L}^*$ are related by 
\begin{figure}
\begin{center}
\subfigure[]{
\centering
\includegraphics[scale=0.16]{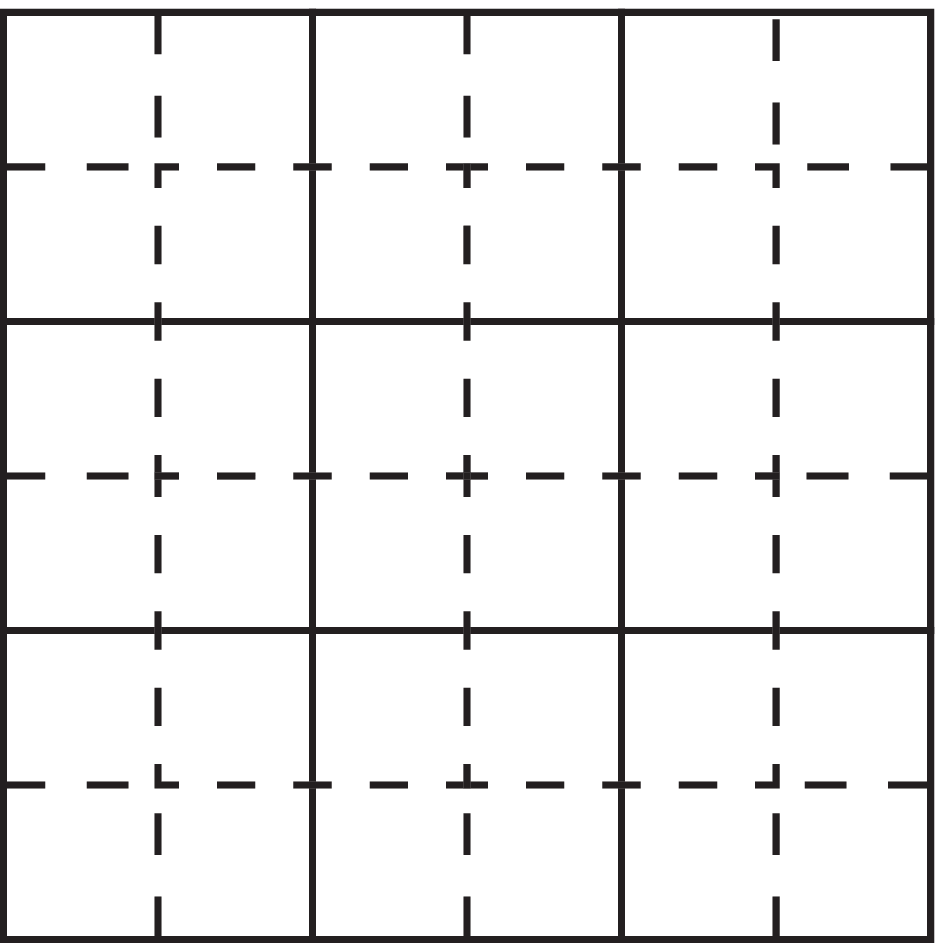}
}
\hspace{3mm}
\subfigure[]{
\centering
\includegraphics[scale=0.16]{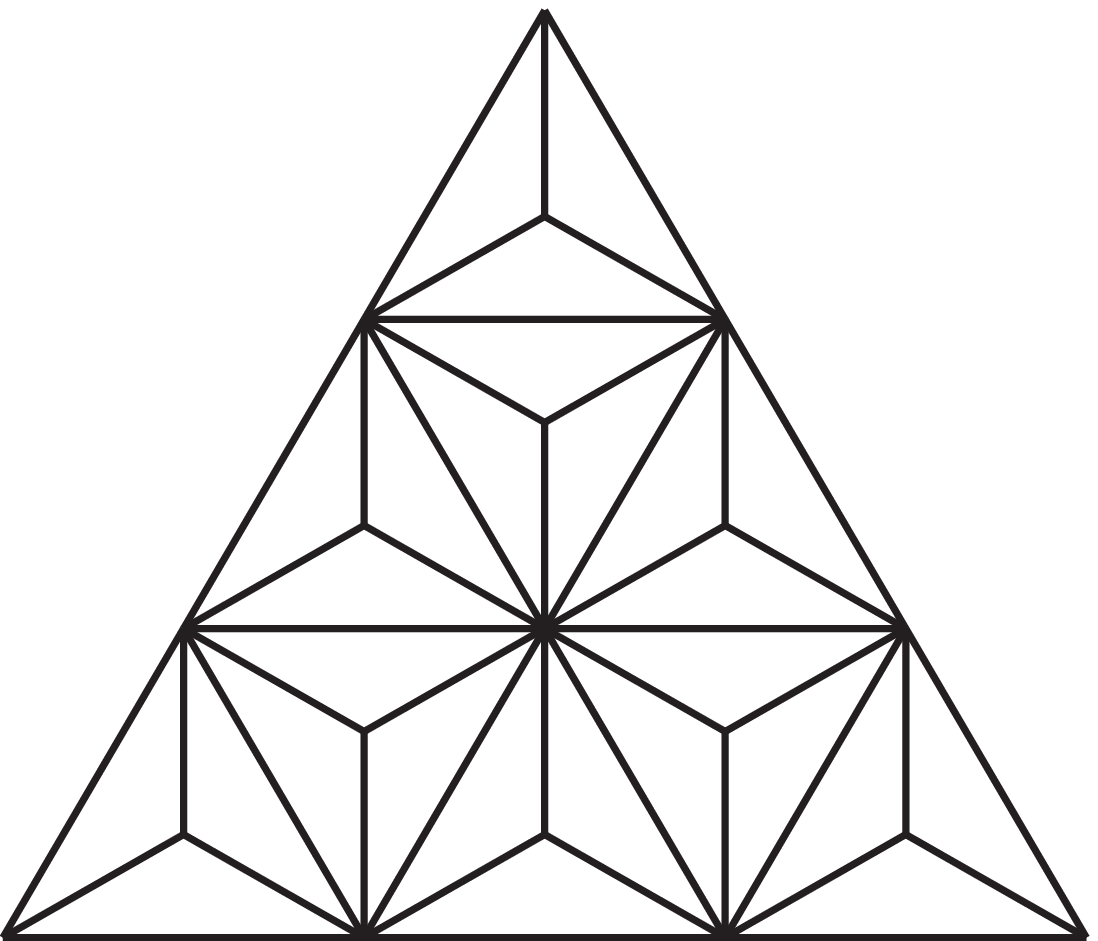}
}
\hspace{3mm}
\subfigure[]{
\centering
\includegraphics[scale=0.16]{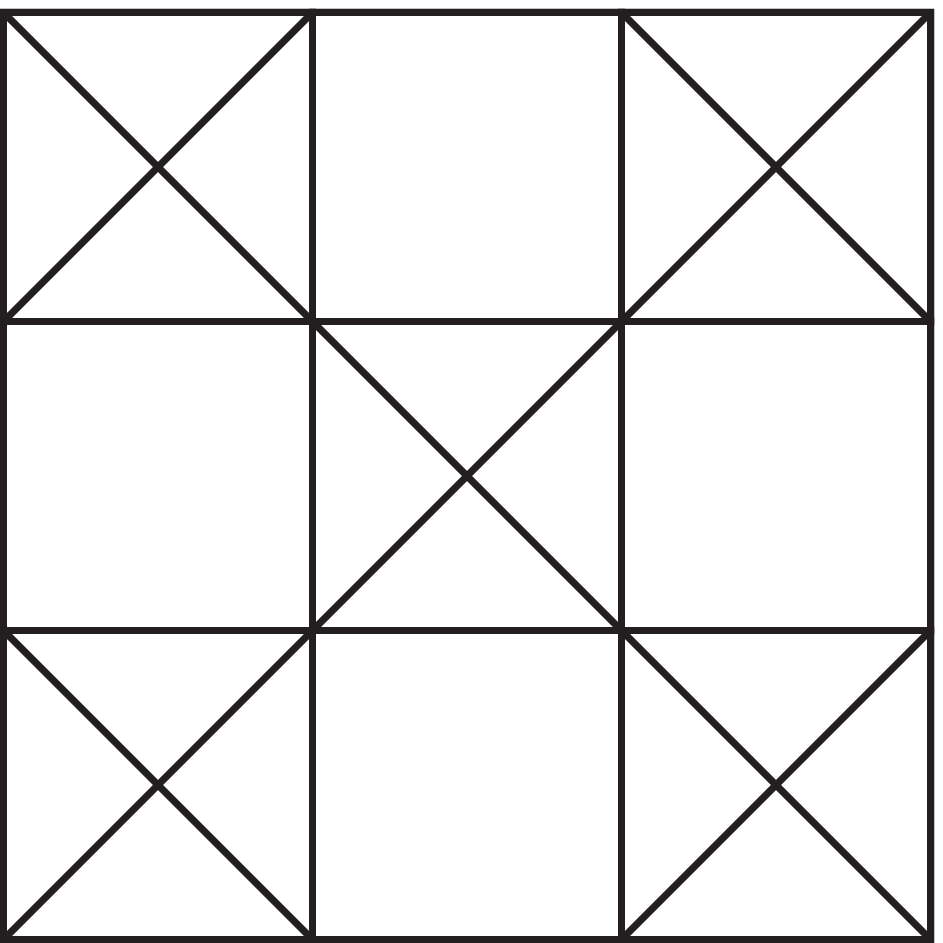}
}
	\caption{(a) A typical example of a self-dual lattice. The square lattice (solid line) and its dual lattice (dash line) are topologically identical. (b) The asanoha lattice [dual to the $(3,12^2)$ lattice]. It is self-matching since any 2D infinite lattice that is fully triangulated is a self-matching lattice. (c) The covering lattice of bond percolation on the square lattice. It is self-matching according to the argument in Ref.~\onlinecite{exact1}. Here the diagonal bonds are nonplanar, and actually for all cases where the faces are not triangular, the matching lattice is always nonplanar.} 
\label{selfex}
\end{center}
\end{figure}
\begin{equation}
p_c^{\textup{bond}}(\mathcal{L})+p_c^{\textup{bond}}(\mathcal{L}^*)=1.
\end{equation}  
Given a planar lattice $\mathcal{L}_0$, a pair of matching lattices can be constructed by doing the following: (i) Select any subset of the faces of $\mathcal{L}_0$, and fill in all the possible diagonals inside these faces to form a new graph $\mathcal{L}$; (ii) Select the faces that are not selected in step (1), and fill in all the possible diagonals in these faces to form another graph $\mathcal{L}^\prime$. For site percolation, a similar relation between a pair of matching lattices $\mathcal{L}$ and $\mathcal{L}^\prime$ is
\begin{equation}
p_c^{\textup{site}}(\mathcal{L})+p_c^{\textup{site}}(\mathcal{L}^\prime)=1.
\end{equation}From Eqs.~$(1)$ and $(2)$, all bond percolation thresholds on the self-dual lattices and site percolation thresholds on the self-matching lattices are known to be $p_c=1/2$. Examples include bond percolation on the square and martini-B lattice, and site percolation on the triangular, union jack, and asanoha [dual to the $(3,12^2)$] lattice~\cite{exact6}. Typical examples of self-dual and self-matching lattices are shown in Fig.~\ref{selfex}.

\begin{figure}
\begin{center}
\subfigure[]{
\centering
\includegraphics[scale=0.15]{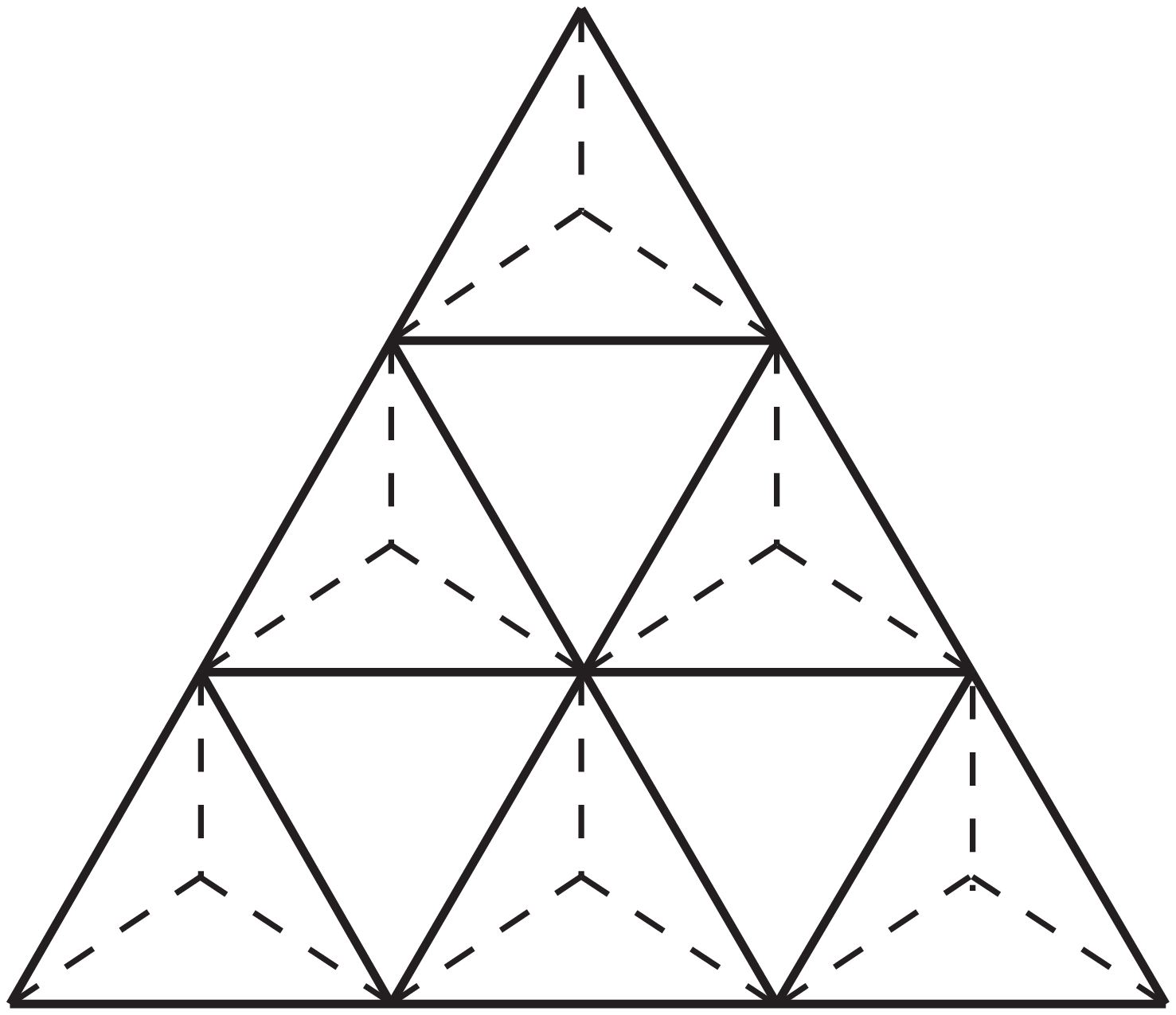}
\label{star1}
}
\hspace{3mm}
\subfigure[]{
\centering
\includegraphics[scale=0.15]{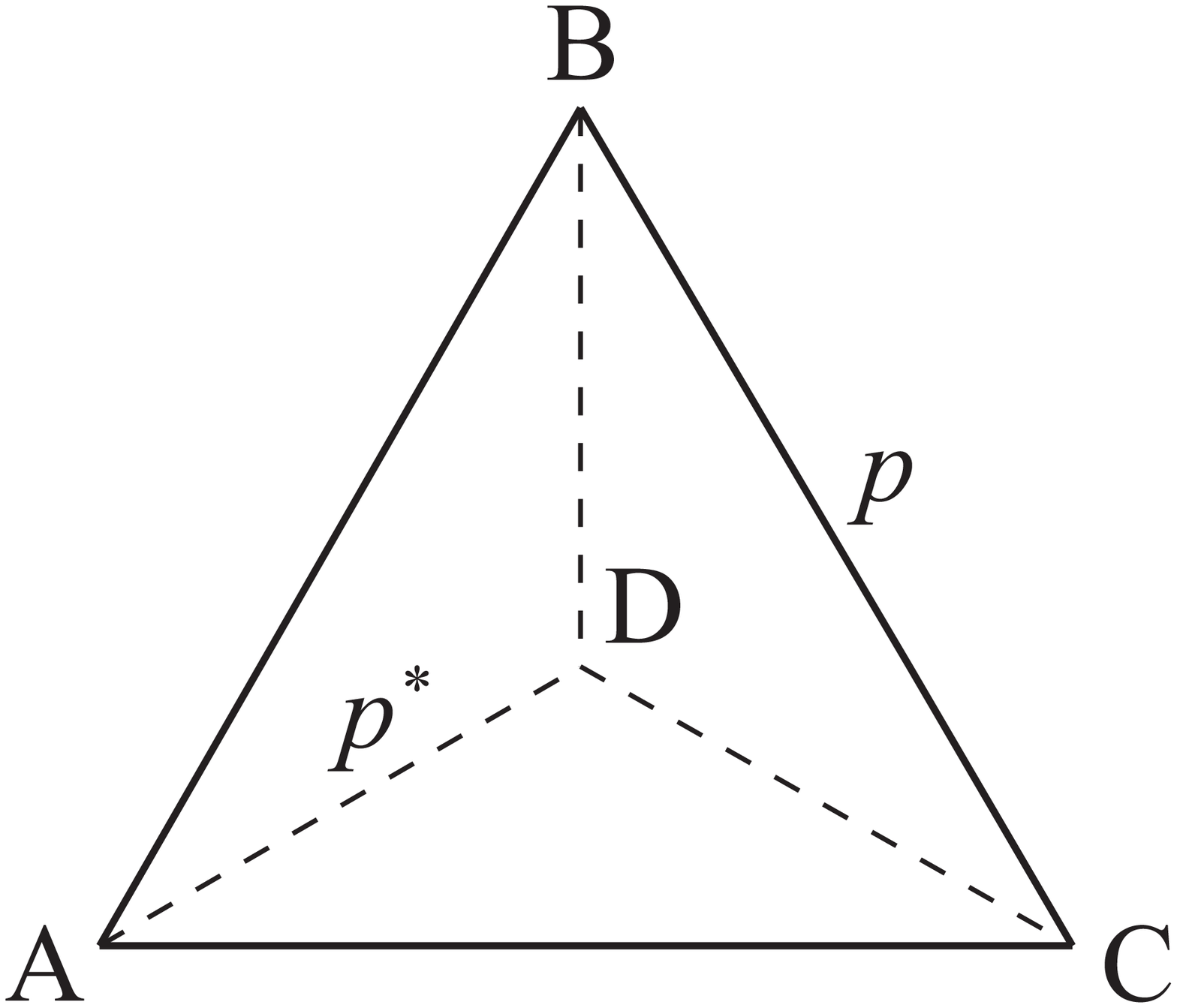}
\label{star2}
}
\caption{(a) The star-triangle transformation on the triangular lattice; (b) One individual star-triangle with the bond probabilities $p$ and $p^*$. } 
\end{center}
\end{figure}

In $1964$, Sykes and Essam~\cite{exact1} introduced into the percolation field the star-triangle transformation, which had been used for electrical circuits~\cite{Kennelly1899} as well as for the Ising model~\cite{Onsager44}. 
By use of the star-triangle transformation and bond-to-site transformation, they found the exact values of bond percolation thresholds on the triangular and honeycomb lattices, and of the site percolation threshold on the kagome lattice. The star-triangle transformation was further generalized for bond percolation on the bowtie lattice in 1984~\cite{exact2} and site percolation on the martini lattice in 2006~\cite{exact3}. Here we simply illustrate this method without proving it. As shown in Fig.~\ref{star1}, one replaces the bonds of every unit cell of the triangular lattice with a star, which transforms the triangular lattice into the honeycomb lattice. Supposing that the bonds of the two lattices are occupied with probabilities $p$ and $p^*$, respectively, and that the corresponding bond thresholds are $p_c$ and $p_c^*$,
one considers bond percolation on an individual ``star-triangle'' shown in Fig.~\ref{star2}. The probability of $A$ being connected to both $B$ and $C$, which is denoted as $P(A \rightarrow B,A \rightarrow C)$ on the triangular lattice and $P^*(A \rightarrow B,A \rightarrow C)$ on the honeycomb lattice, can be obtained as
\begin{equation*}
P(A \rightarrow B,A \rightarrow C)=3p^2-2p^3
\label{stareq1}
\end{equation*}
and
\begin{equation*} 
P^*(A \rightarrow B,A \rightarrow C)={p^*}^3.
\label{stareq2}
\end{equation*}
Following the argument in Ref.~\onlinecite{exact1}, the critical surface is defined as 
\begin{equation}
P(A \rightarrow B,A \rightarrow C)=P^*(A \rightarrow B,A \rightarrow C).
\label{stareq3}
\end{equation}
Moreover, the duality between the triangular and honeycomb lattices guarantees that $p_c$ and $p_c^*$ are related by Eq.~$(1)$. Combining Eq.~$(1)$ and $(3)$, one obtains
\begin{equation}
p_c^3-3p_c+1=0.
\end{equation} Eq.~$(4)$ has only one root at $p_c=2 \sin{\pi/18}$ in the range $[0,1]$, which is exactly the bond percolation threshold of the triangular lattice. Besides Eq.~$(3)$, there are other connectivities that should be tested. For example, the probability of $A$ being connected to $B$ but not $C$, denoted $P(A \rightarrow B,A \nrightarrow C)$, is
\begin{equation*}
P(A \rightarrow B,A \nrightarrow C)=p(1-p)^2
\label{stareq1}
\end{equation*}and 
\begin{equation*}
P^*(A \rightarrow B,A \nrightarrow C)=(1-p^*){p^*}^2.
\label{stareq1}
\end{equation*}It is noted that $P(A \rightarrow B,A \nrightarrow C)=P^*(A \rightarrow B,A \nrightarrow C)$ leads to Eq.~$(1)$. Thus one cannot obtain an additional relation from the former equation, and it is similar for $(A \nrightarrow B,A \rightarrow C)$ and $(B \rightarrow C,B \nrightarrow A)$ cases. The condition $P(A \nrightarrow B,A \nrightarrow C)=P^*(A \nrightarrow B,A \nrightarrow C)$, however, is equivalent to Eq.~$(3)$. Generally speaking, the connectivity probabilities on both ``star" and ``triangle" are required to be equivalent at criticality.

In 2006, Scullard and Ziff~\cite{Ziff06, exact5} introduced the triangle-triangle transformation. This method extends the star-triangle transformation to lattices in which the basic cells do not necessarily lie in a triangular lattice, but in any self-dual arrangement. 
\begin{figure}
\begin{center}
\subfigure[]{
\includegraphics[scale=0.2]{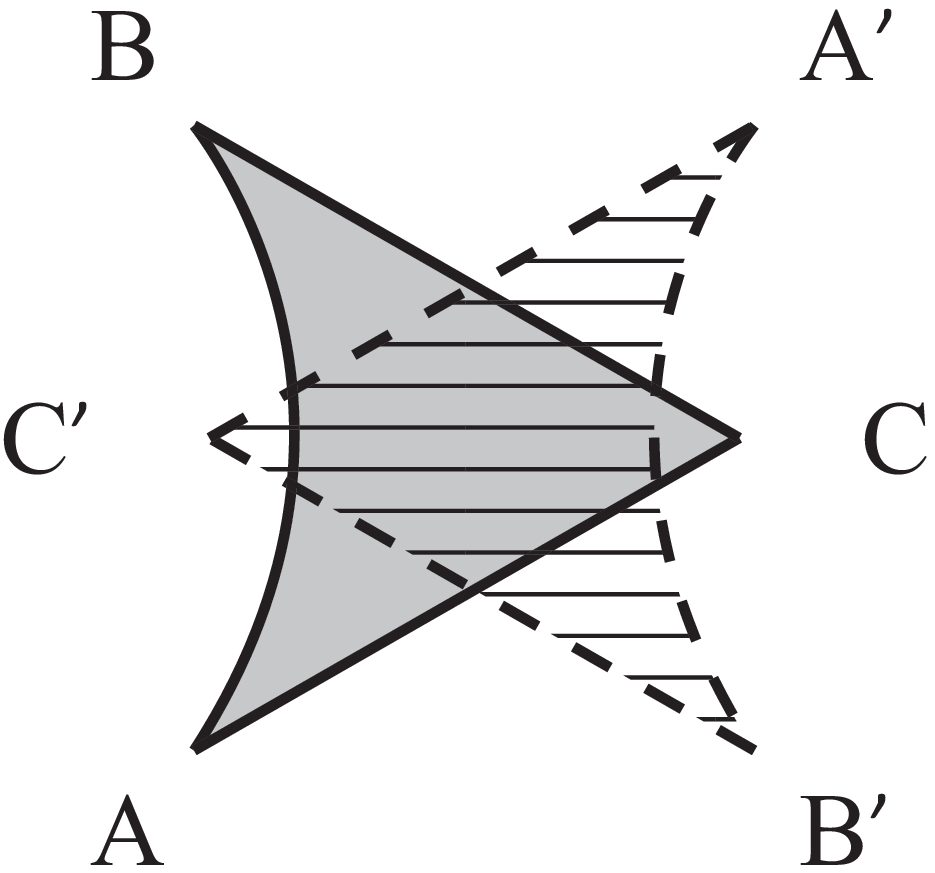}
}
\\
\subfigure[]{
\includegraphics[scale=0.09]{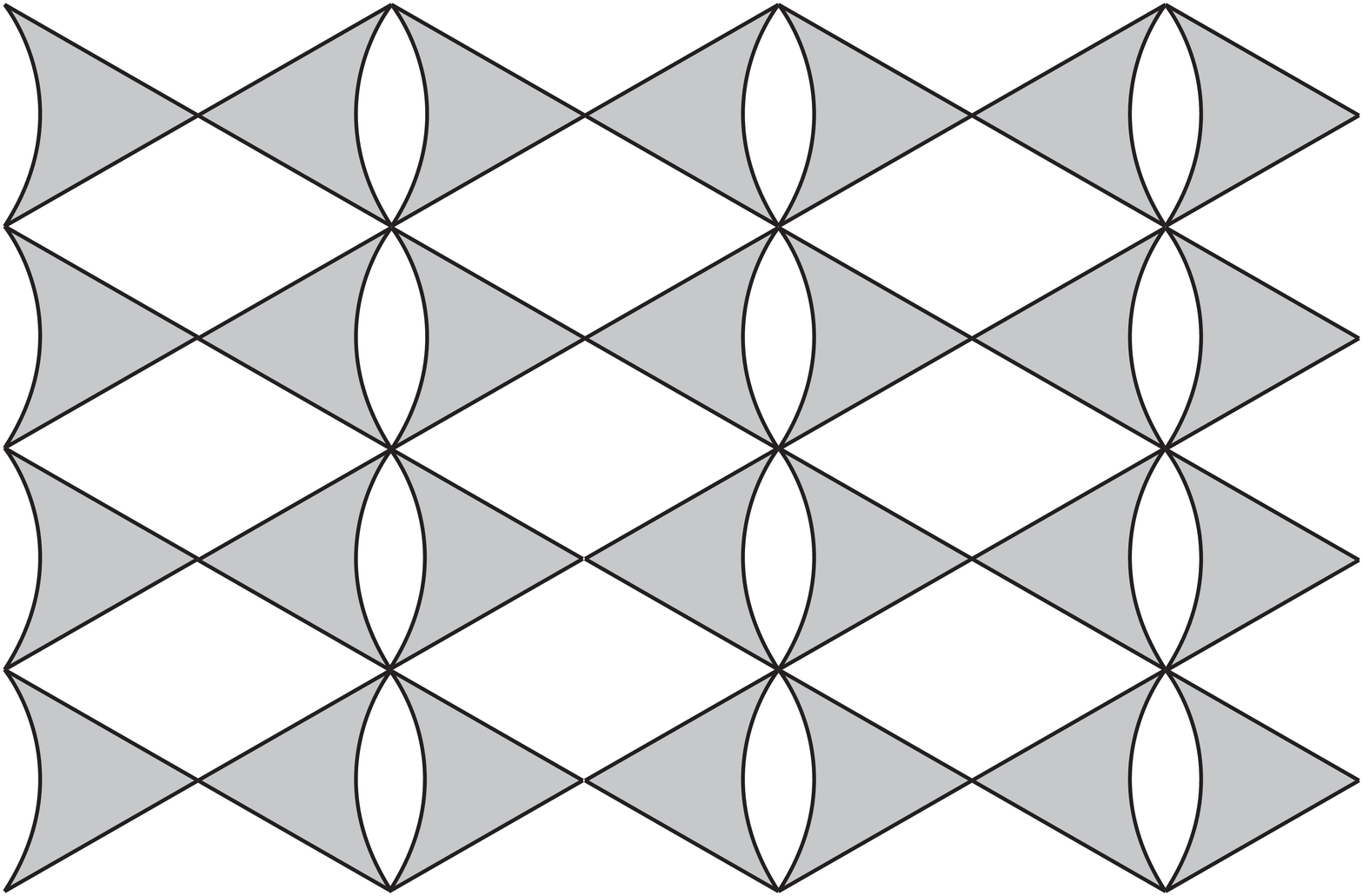}
}
\hspace{3mm}
\subfigure[]{
\includegraphics[scale=0.09]{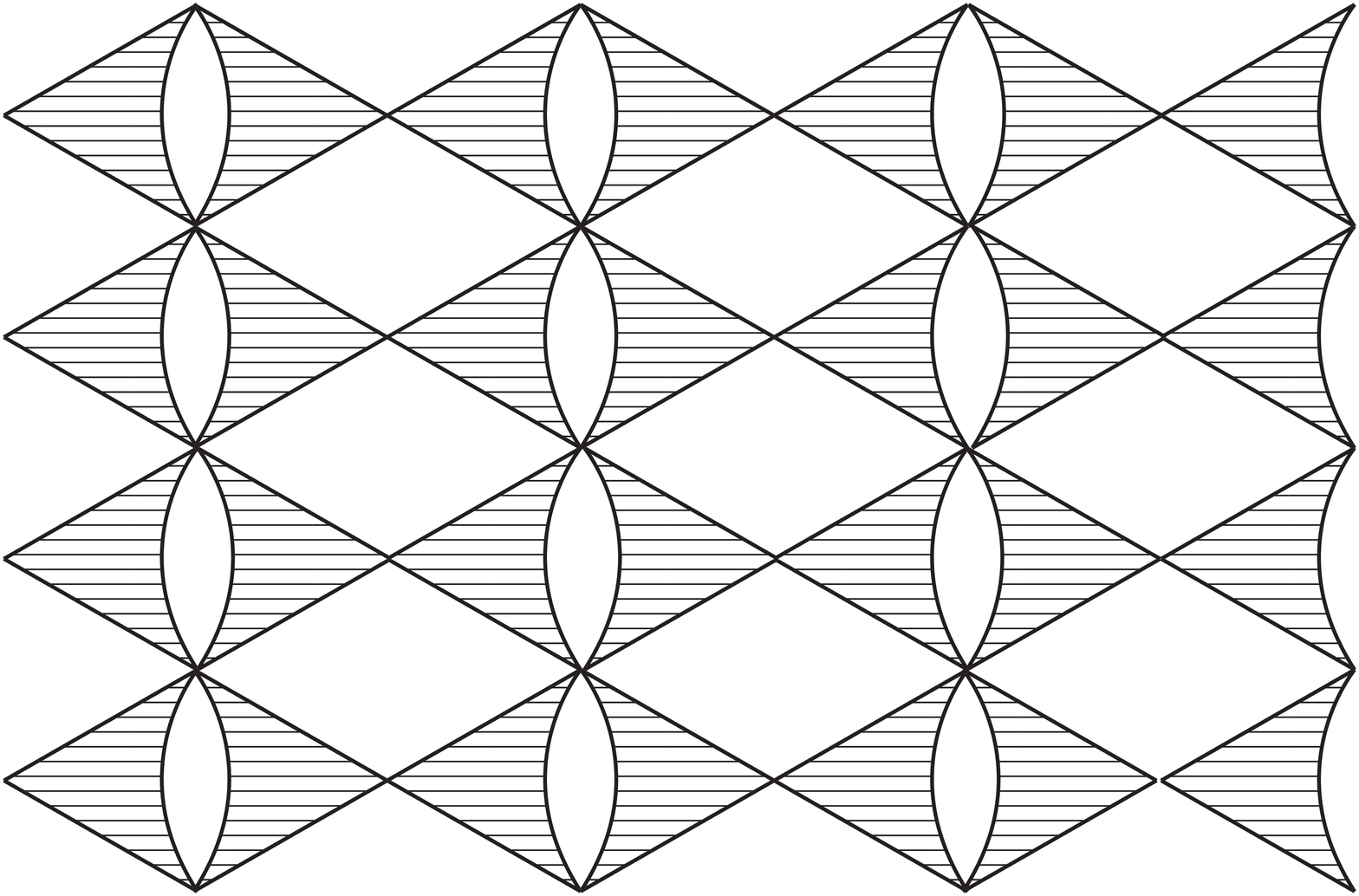}
}

	\caption{The triangle-triangle transformation on a basic cell is shown in (a). The shaded region can contain any interactions among the vertices A, B, C. (b) is an example of a self-dual lattice (the bowtie graph) since it is invariant under this transformation, as shown in (c).}
\label{tri-tri}
\end{center}
\end{figure}
Here a ``self-dual" lattice is defined as a lattice which is invariant under the triangle-triangle transformation, as shown in Fig.~\ref{tri-tri}. The basic cell can represent any network of bonds and sites contained within the vertices $A$, $B$, $C$, as long as no sites are at these vertices. Similarly, they consider the connectivity between the vertices, which yields a general condition for criticality as
\begin{equation}
	P_{\Delta}(A,B,C)=P_{\Delta}(\bar{A},\bar{B},\bar{C}) \,,
\end{equation} 
where $P_{\Delta}(A,B,C)$ refers to the probability that three vertices $A$, $B$, $C$ are connected, and $P_{\Delta}(\bar{A},\bar{B},\bar{C})$ refers to the probability that none are connected. 
Equation~$(5)$ leads to the threshold for any lattice that is self-dual under triangle-triangle transformation, and therefore significantly expands the number and types of lattices with exactly known thresholds~\cite{exact5,Wierman11}. 
For example, one can apply Eq.~$(5)$ to get bond percolation thresholds of the square, triangular and honeycomb lattices. Other examples include site and bond percolation thresholds for the ``martini", ``martini-A", ``martini-B" and bowtie lattices~\cite{exact5,Wierman11}.
The approach is also applied to determine the critical manifolds of inhomogeneous bond percolation on bowtie and checkerboard lattices~\cite{Ziff12}, although for the latter and some cases of the former one needs to introduce artificial bonds with negative probability. 
It is noted that for the checkerboard case, the approach reproduces F. Y. Wu's formula~\cite{Wu79}, which can be proven by the isoradial construction~\cite{Ziff12,Grimmett14,Kenyon04}.

In the past few years, Scullard, Ziff and Jacobsen developed the so-called critical polynomial method~\cite{origin,gen0,gen2,gen1,j2,j6,j4,j5,j1,j3} which associates a graph polynomial with any two dimensional (2D) periodic lattice. This method originates from the observation that all the exact percolation thresholds appear as the roots of polynomials with integer coefficients. For example, the bond threshold of the triangular lattice is the root of integer polynomial shown in Eq.~$(4)$. Scullard and Ziff first defined such a polynomial based on the linearity hypothesis and symmetries~\cite{origin,gen0}. By employing a deletion-contraction algorithm, this polynomial can be applied on any 2D periodic lattice and provide, in principle, arbitrarily precise approximations for percolation thresholds~\cite{gen2,gen1}. Scullard and Jacobsen further gave an alternative probabilistic definition of the critical polynomial~\cite{j2,j6} which allows for much more efficient computations~\cite{j4,j5,j1,j3}.

\begin{figure}
\begin{center}

\includegraphics[scale=0.18]{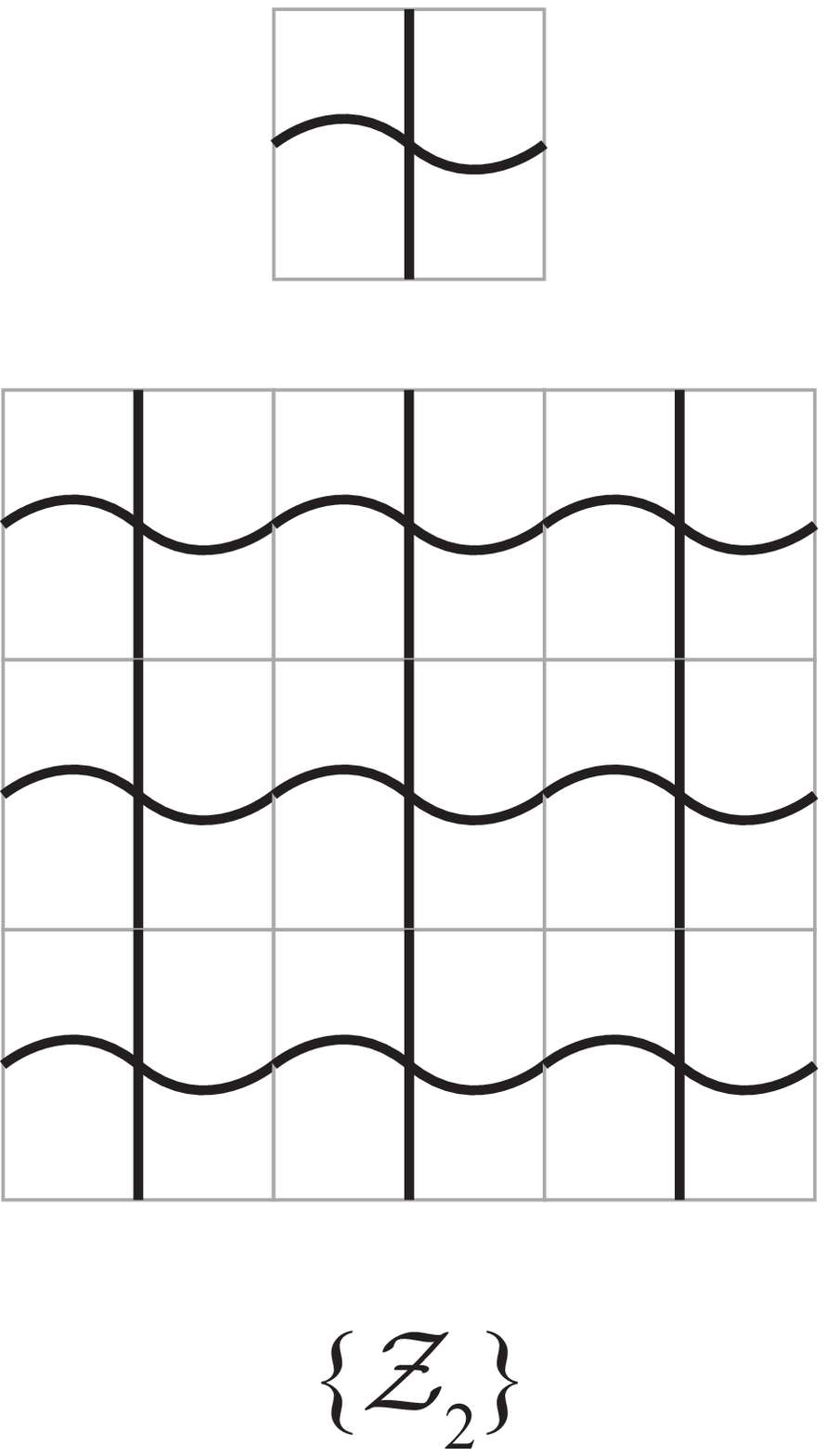}
\hspace{2mm}
\includegraphics[scale=0.18]{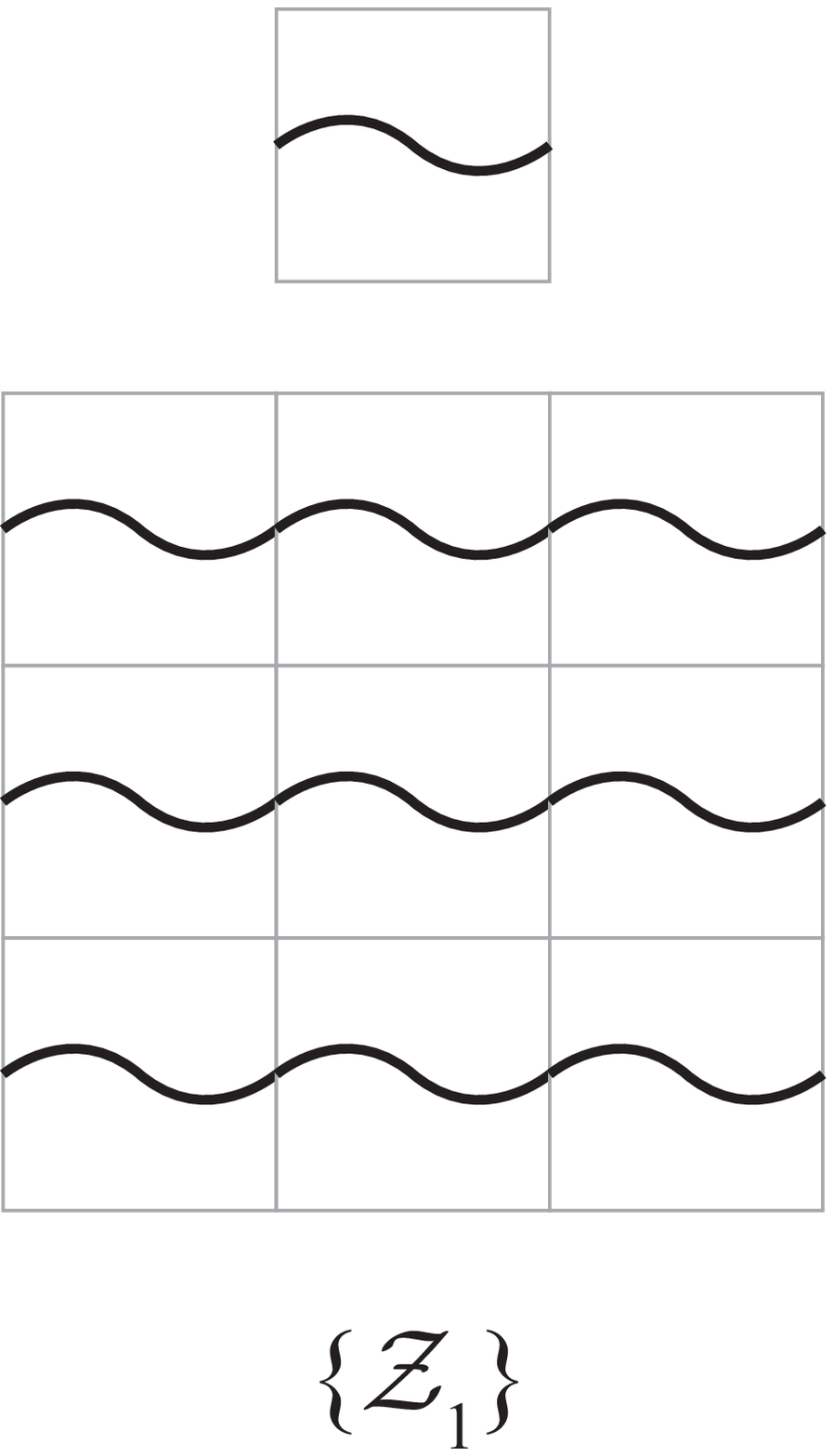}
\hspace{2mm}
\includegraphics[scale=0.18]{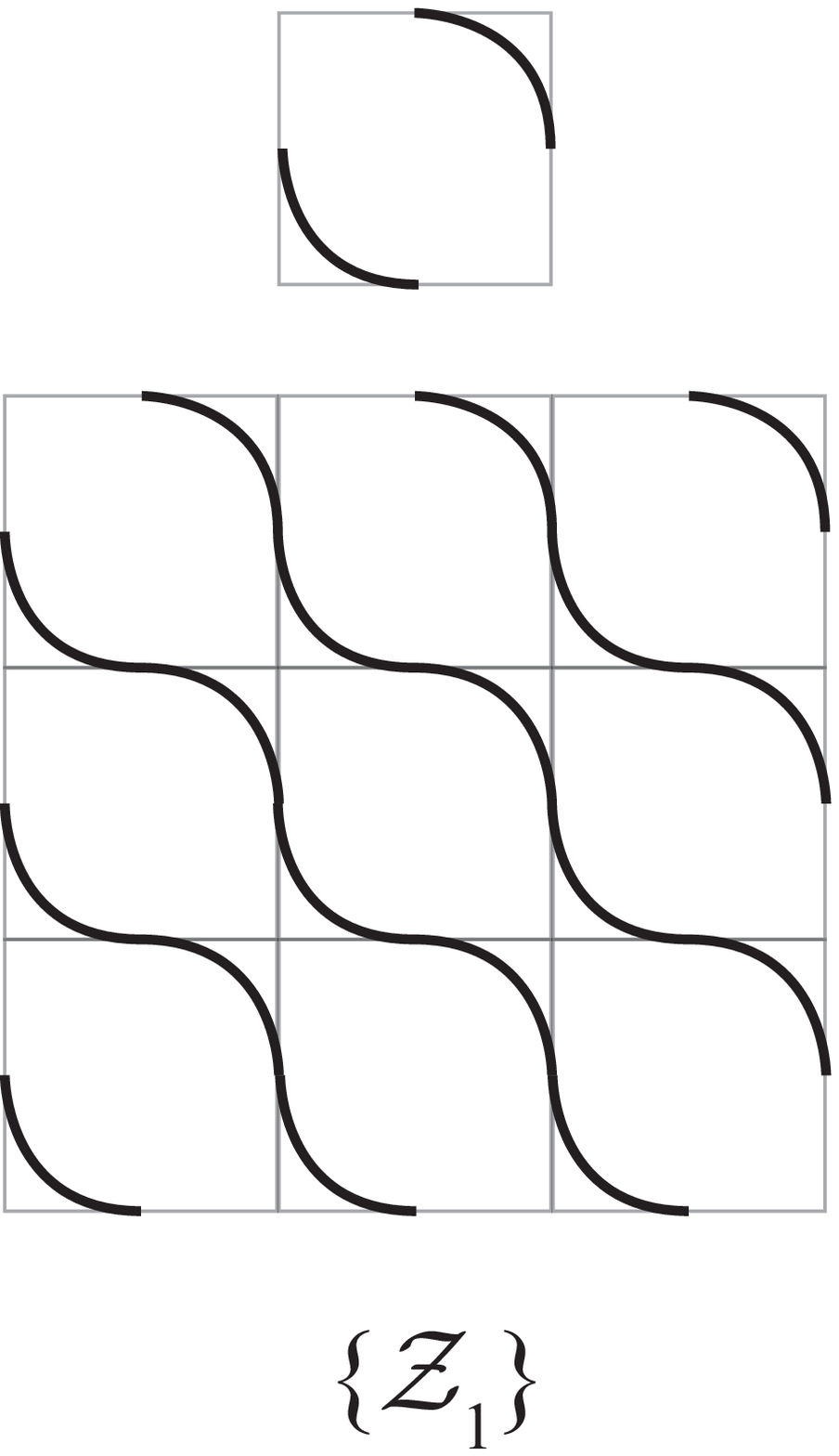}
\hspace{2mm}
\includegraphics[scale=0.18]{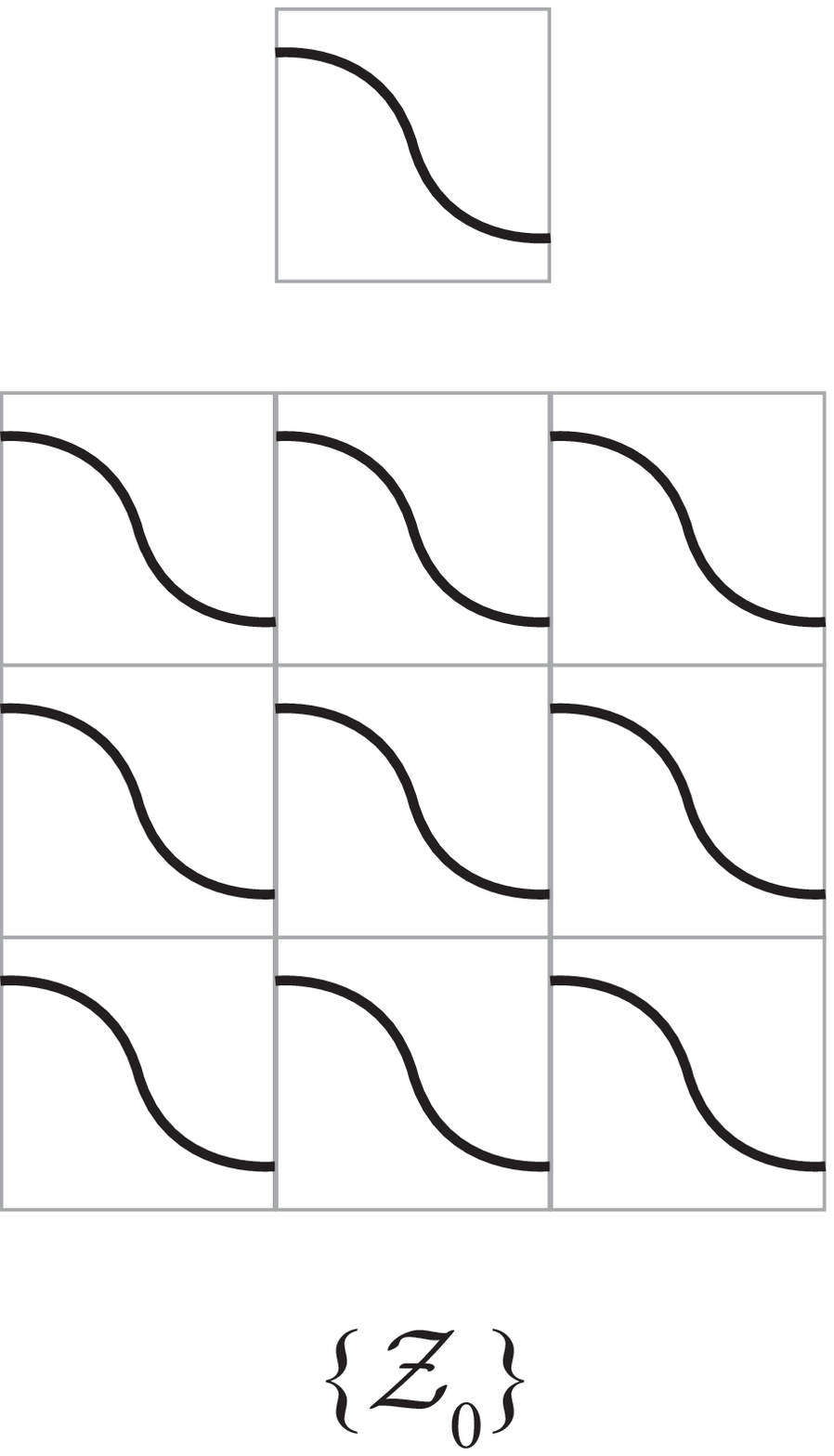}

\caption{Typical configurations on a 2D square with periodic boundary conditions. The number of different directions along which a configuration wraps decides its type among $\{\mathcal{Z}_2\}$, $\{\mathcal{Z}_1\}$, and $\{\mathcal{Z}_0\}$.}
\label{cp}
\end{center}
\end{figure}

For simplicity, we describe the critical polynomial on a 2D square with periodic boundary conditions (a torus). All the configurations $\{\mathcal{C}\}$ on the torus are classified into three types as $\{\mathcal{Z}_0\}$, $\{\mathcal{Z}_1\}$, and $\{\mathcal{Z}_2\}$ according to their topological properties. 
As shown in Fig.~\ref{cp}, a configuration $\mathcal{C}$ belongs to $\{\mathcal{Z}_2\}$ if it wraps along two different directions, to $\{\mathcal{Z}_1\}$ if it wraps along one and only one direction, and to $\{\mathcal{Z}_0\}$ if it does not wrap. $R_2$, $R_1$ and $R_0$ represent the probabilities for a configuration to be in these classes respectively, i.e., the wrapping probabilities~\cite{wrap1,wrap2,wrap3}. 
For planar lattices, when the configuration is of $\mathcal{Z}_2$-type ($\mathcal{Z}_0$-type), the corresponding configuration on the dual lattice is of $\mathcal{Z}_0$-type ($\mathcal{Z}_2$-type). This duality relation leads to $R_2 = R_0$ for self-dual lattices at critical point. Wrapping probabilities $R_2$ and $R_0$ are polynomial functions of the occupation probability $p$, and generally the critical polynomial is defined as $\PB \equiv R_2-R_0$. 
From universality of $R_2$ and $R_0$, the condition for criticality can be written as
\begin{equation}
\PB(p,L)=0.
\end{equation}
The properties of $\PB$ on planar lattices are as follows:  
\begin{itemize}
    \item The root of Eq.~$(6)$ provides an estimate for percolation the threshold $p_c$, and it satisfies $\lim\limits_{L \to \infty}p(L)=p_c$.
    \item Finite-size correction vanishes for all solvable lattices: $\PB(p_c,L)=0$. Therefore, the root of Eq.~$(6)$ gives the exact value of $p_c$ for arbitrary system size $L$.
    \item $(p(L)-p_c) \simeq \sum_{k=1}^{\infty}{A_kL^{- \Delta_k}}$ vanishes rapidly for those lattices of which the $p_c$ value is not exactly known. For unsolved Archimedean lattices, it is suggested that there are two different classes: one has the first three scaling exponents $\Delta=6,7,8$, and the other has $\Delta=4,6,8$~\cite{j3}.
  \end{itemize}

\begin{figure}
\begin{center}

\subfigure[]{
\includegraphics[scale=0.18]{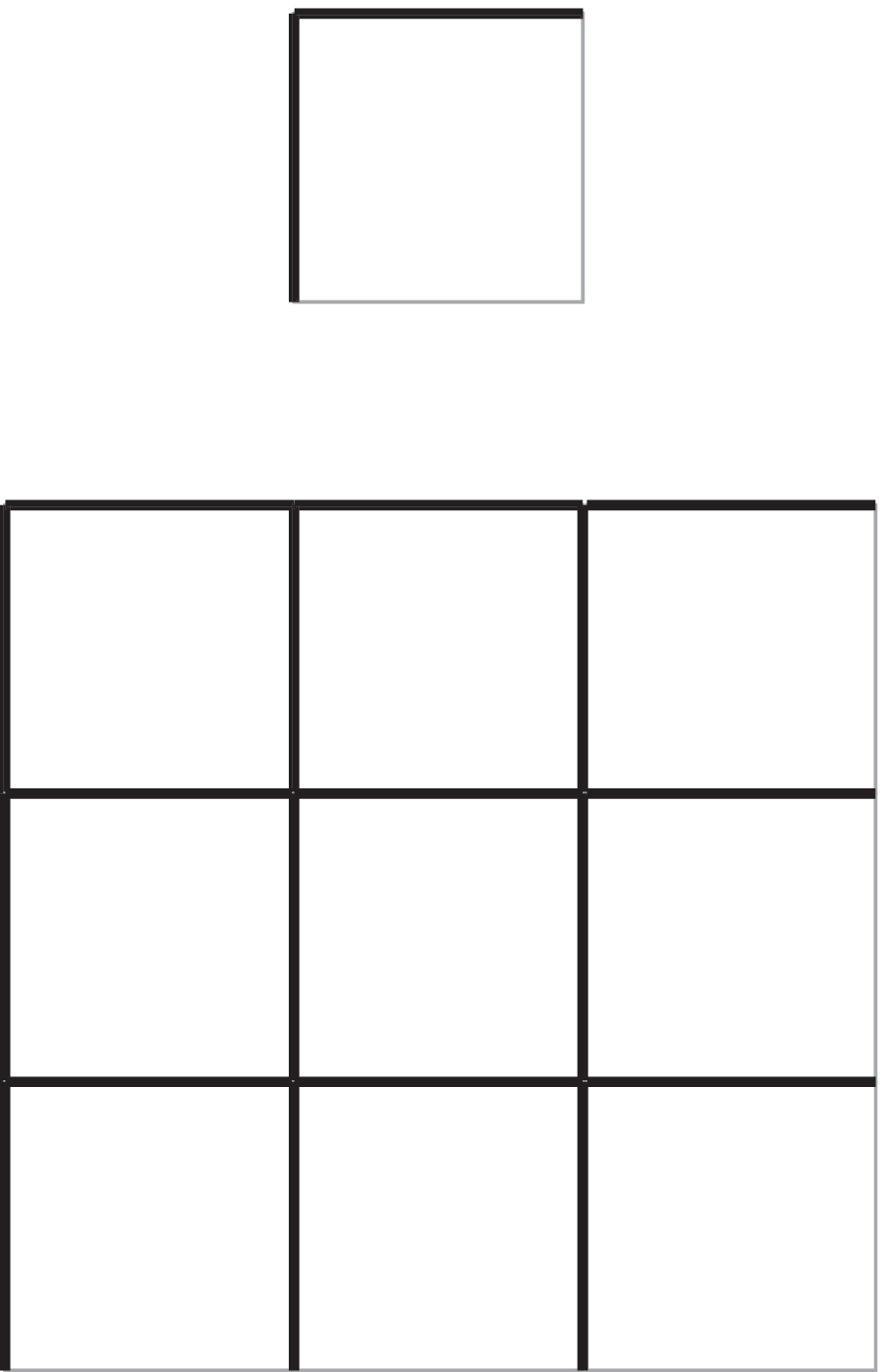}
\label{cpe}
}
\hspace{5mm}
\subfigure[]{
\includegraphics[scale=0.18]{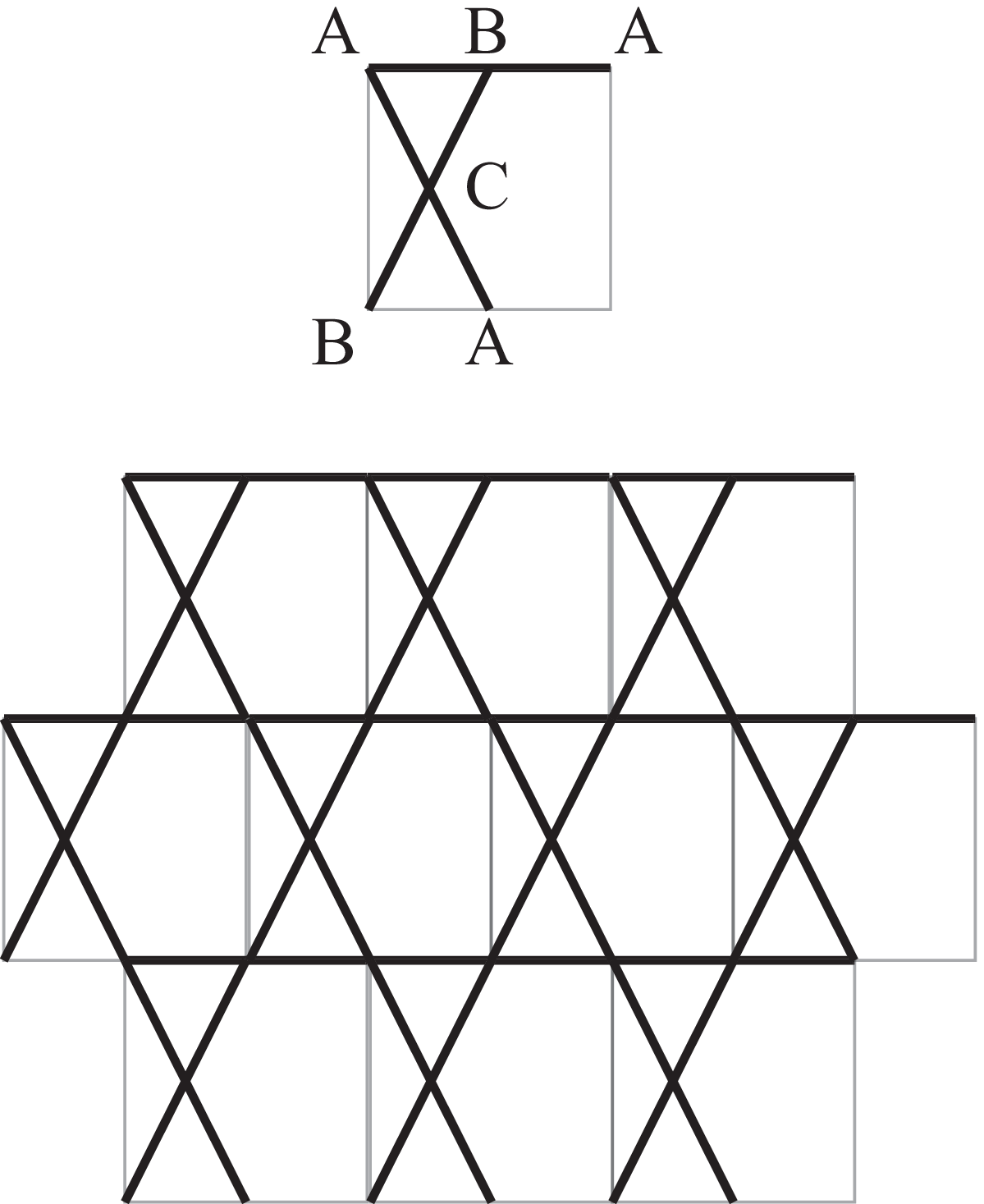}
\label{cpe2}
}

\caption{(a) The basic cell of the square lattice with the bond probability $p$. (b) The basic cell of the kagome lattice with the site probability $q$.}
\end{center}
\end{figure}

Here we further explain these properties. For solvable lattices, the root of Eq.~$(6)$ in $[0,1]$ agrees with the exactly known thresholds regardless of the system size. A simple example is bond percolation on the square lattice. Consider the smallest repeated cell of the square lattice as shown in Fig.~\ref{cpe}, and suppose each bond is occupied independently with probability $p$. The wrapping probabilities can be easily calculated as $R_2=p^2$ and $R_0=(1-p)^2$, and therefore $\PB=p^2-(1-p)^2$. The only root of Eq.~$(6)$ is $p=1/2$ which is exactly the bond percolation threshold of the square lattice. Another example is site percolation on the kagome lattice as shown in Fig.~\ref{cpe2}. 
The basic cell contains three vertices $A$, $B$, $C$ that are independently occupied by sites with probability $p$, which is different from the cell in Fig.~\ref{tri-tri}(a) for the triangle-triangle transformation where the vertices are not allowed to be occupied by sites.
We calculate the wrapping probabilities as $R_2=p^3$ and $R_0=3p(1-p)^2+(1-p)^3$, which lead to 
$\PB=(1-p)^3-3(1-p)+1$. 
Thus the site percolation threshold of the kagome lattice is given by the root of Eq.~$(6)$ as $p_c=1-2 \sin{\pi/18}$, which is identical with the bond percolation threshold of the honeycomb lattice.  
This is a natural result because site percolation on the kagome lattice is isomorphic with bond percolation on the honeycomb lattice according to the bond-to-site transformation.

For many unsolved 2D periodic lattices, the critical polynomial method has been shown to be orders of magnitude more accurate in determining the percolation threshold than traditional techniques, because of its surprisingly small finite-size corrections. It has also been applied to the $q$-state Potts model in the Fortuin-Kasteleyn representation to predict critical manifolds~\cite{j6,j4,j1} with $\PB \equiv R_2-qR_0$, where $q$ is related to the symmetry of the model and $q \rightarrow 1$ corresponds to percolation. The generalization to nonplanar and continuum models, as far as we know, has not been reported yet. In these models, the value of $\PB$ in the scaling limit is supposed to be zero as well due to universality, but the finite-size scaling (FSS) behavior is not clear.

The goal of this work is to explore the FSS behavior of $\PB$ in nonplanar and continuum systems. 
For comparison purpose,  FSS analysis is also performed for the wrapping probability $R_2$ 
and a dimensionless ratio $Q$ related to the size of the largest cluster.
Extensive Monte Carlo (MC) simulations are 
conducted for a nonplanar lattice model, i.e., the 2D square-lattice bond percolation with many equivalent neighbors~\cite{Ouyang18, Deng19},
and for the 2D continuum percolation with identical penetrable disks~\cite{MM}.
Periodic boundary conditions are employed as required for measuring $\PB$.
The simulation results confirm that $\PB=0$ for these two models at the critical point. 

For the equivalent-neighbor percolation model, one of us (YD) and collaborators~\cite{Ouyang18,Deng19} observed recently that
as long as the coordination number $z$ is finite, the model belongs to the short-range universality in two dimensions. The percolation threshold was determined by the critical polynomial, but the analysis details have not been reported. It is particularly informative to compare the finite-size correction in $P_{\rm B}$ and  in more conventional quantities. 
In this work, the finite-size correction in $\PB$ is found 
to be very small. 
For the model with $z=8$ equivalent neighbors,
the leading correction term of $\PB$ scales as $L^{y_1}$ with $y_1 \simeq -3$, 
while for $R_2$ and $Q$ the leading correction term is of order $L^{-2}$ or larger.
For $z>8$, two types of models are considered,
which have different ways to involve neighbors, 
i.e., by coupling to all sites within a circle or a square.  
It is shown that the data of $\PB$ are still consistent with the leading correction term 
being $b_1 L^{-3}$. However, the amplitude $b_1$ cannot be well determined 
by fitting the data, which indicates that our data are barely sufficient to detect 
the small finite-size correction.
For very large $z$, e.g., $z \sim {\text{O}}(10^5)$, due to finite-size corrections,
for sizes up to $L=8192$, the crossing points of the wrapping probability 
deviate significantly from the percolation threshold, 
and the dimensionless ratio does not show a crossing at all in a wide range near $p_c$.
Thus it is very hard to use the wrapping probability or the dimensionless ratio
to determine precisely the percolation threshold for large $z$, as simulations for 
much larger $L$ are needed.
By fitting the FSS ansatz of $\PB$, 
it is possible to determine precisely values of $z p_c$ for $z$ up to O$(10^5)$~\cite{Deng19}. 
The data confirm the $z \rightarrow \infty$ asymptotic behavior $z p_c -1 \simeq a_1 z^{-1/2}$ 
for both types of models, and show that the coefficient $a_1$ 
takes different values for the two models. The latter indicates that $a_1 z^{-1/2}$ 
represents a surface effect for the 2D model~\cite{Frei16, Lalley14}. 

For the continuum model, it is found that at criticality the finite-size correction in $\PB$ is too small to be observed for $L\ge3$, i.e., $\PB(\rho_c,L)=0$ almost holds for arbitrary $L$. 
In comparison, a leading correction term $\sim L^{-2}$ is confirmed for $R_2$ and $\sim L^{-1.5}$ for $Q$. Using $\PB$, the percolation threshold of the continuum model is determined as $\rho_c = 1.436\,325\,05(10)$, slightly below the most recent result $\rho_c = 1.436\,325\,45(8)$ given by Mertens and Moore~\cite{MM}. 

The remainder of this work is organized as follows. 
Sec.~\ref{sec-equi} presents the simulation and results for 
the square-lattice bond percolation model with various number of equivalent neighbors, 
and Sec.~\ref{sec-cont} describes those for the 2D continuum percolation model. 
A brief discussion and conclusion is given in Sec.~\ref{sec-dis}.

\section{Equivalent-neighbor percolation}
\label{sec-equi}

\subsection{Model and simulation}
To the best of our knowledge, the equivalent neighbor model was first
introduced by Domb and Dalton~\cite{Domb66, Dalton66} 
to help bridge the gap in the understanding of spin systems 
between very short-range forces and very long-range forces.
Recently, equivalent-neighbor percolation models were studied
for bond percolation in 2D~\cite{Ouyang18, Deng19}, 3D~\cite{Xun20}
and 4D~\cite{Xun20b}, and for site percolation in 2D~\cite{Malarz07, Koza14, XunHaoZiff20}
and 3D~\cite{Malarz15, XunHaoZiff20}.
In the square-lattice bond percolation model with equivalent neighbors,
for each lattice site, there exists an edge between
this site and any site within a given range. 
Two sites at the end of the same edge are called neighbors. 
Two ways to involve neighbors are considered:
in type-1 model a site $i$ with coordinates $(x_i,y_i)$ is connected by an edge
to all sites $j$ satisfying $\sqrt{(x_i-x_j)^2 + (y_i-y_j)^2} \le r$ (i.e., within
a circle of radius $r$),
and in type-2 model to all sites $j$ satisfying both $|x_i-x_j| \le r$ 
and $|y_i-y_j| \le r$ (i.e., within a square of side length $2r$).
Similar to the nearest-neighbor percolation, the equivalent-neighbor percolation
is introduced by placing independently a bond on each edge with the same probability $p$. 

We simulate the above models with periodic boundary conditions. 
Since there are many equivalent neighbors, the simulation 
would be time consuming if the edges are individually checked
to be occupied or not. We apply an algorithm~\cite{Luijten95, Deng19} 
which requires computer time that is almost independent of the number of neighbors $z$.
The cluster wrapping is detected by a method~\cite{Machta96, wrap4} 
originally employed in simulations of Potts models.
Quantities are sampled after all the clusters are constructed 
and a configuration is formed. For all the configurations, 
the following observables are sampled:
\begin{itemize}
    \item The critical polynomial $\PB$ and wrapping probabilities $R_0$, $R_1$ and $R_2$.
    \item The size of the largest cluster $\mathcal{C}_1$. 
    \item The dimensionless ratio $Q={\langle {\mathcal{C}_1}\rangle}^2/{\langle{{\mathcal{C}_1}}^2\rangle}$.
\end{itemize}

Simulations were first performed 
for the model with $z=8$ neighbors. The type of the model is not specified, 
since the type-1 model shares the same $8$ neighbors with the type-2 model.
The system sizes in simulations range from $L=4$ to $64$, 
and the number of samples for each size at a given $p$ 
is around $10^{10}$ to $10^{11}$.  
Simulations were also conducted for several values of $z$ from $148$ ($r=7$) 
to $50616$ ($r=127$) for the type-1 model, 
and from $120$ ($r=5$) to $65024$ ($r=127$) for the type-2 model.
The system sizes for these models of $z>8$ range from $L=16$ to $8192$.

\subsection{Numerical results}

\begin{table*}[htbp]
\begin{center}
\caption{Fit results of the critical polynomial $\PB$ for bond percolation 
	on the square lattice with $z=8$ equivalent neighbors.
	Entries ``--" indicate that the corresponding parameters are set to be zero, 
	and the numbers without error bars are fixed in the fits.} 
\label{Tab:fit-PB}
\begin{tabular}[t]{l|l|l|l|l|l|l|l|l|l}
\hline
	$L_{\rm min}$   &  $\chi^2/$DF  & $y_t$        & $p_c$                & $P_{{\rm B}0}$      & $q_1$           & $b_1$           & $y_1$        & $b_2$       & $y_2$      \\
\hline
   8               &  25.8/32      & 0.84(9)      & 0.250\,368\,50(7)    & 0.000\,008(5) & $-2.7(8)$       & $-0.17(2)$      & $-2.98(7)$   &~~~--        &~~~--        \\
   9               &  23.8/27      & 0.82(9)      & 0.250\,368\,50(8)    & 0.000\,007(6) & $-3(1)$         & $-0.18(4)$      & $-3.0(1)$    &~~~--        &~~~--        \\
   5               &  34.0/38      & 0.84(8)      & 0.250\,368\,50(7)    & 0.000\,007(5) & $-2.7(8)$       & $-0.34(10)$     & $-3.19(10)$  & $0.5(2)$    & $-4$        \\
   6               &  31.4/37      & 0.84(8)      & 0.250\,368\,46(7)    & 0.000\,004(5) & $-2.7(8)$       & $-0.30(9)$      & $-3.2(2)$    & $1.5(6)$    & $-5$        \\
   6               &  30.8/37      & 0.84(8)      & 0.250\,368\,46(7)    & 0.000\,004(5) & $-2.7(8)$       & $-0.24(5)$      & $-3.12(10)$  & $5(2)$      & $-6$        \\
   8               &  30.6/34      & 3/4         & 0.250\,368\,40(2)    & 0             & $-3.6(2)$       & $-0.21(2)$      & $-3.08(4)$   &~~~--        &~~~--        \\
   9               &  26.4/29      & 3/4         & 0.250\,368\,40(2)    & 0             & $-3.7(2)$       & $-0.23(3)$      & $-3.14(6)$   &~~~--        &~~~--        \\
  10               &  29.8/29      & 3/4         & 0.250\,368\,39(2)    & 0             & $-3.7(2)$       & $-0.169(2)$     & $-3$         &~~~--        &~~~--        \\
  12               &  24.0/24      & 3/4         & 0.250\,368\,39(2)    & 0             & $-3.7(2)$       & $-0.165(3)$     & $-3$         &~~~--        &~~~--        \\
\hline
\end{tabular}
\end{center}
\end{table*}

The data of $\PB$ are fitted by the least-square criterion 
using the following ansatz
\begin{equation}
O(p,L)=O_0+q_1(p_c-p)L^{y_t}+b_1L^{y_1}+b_2L^{y_2} \,,
\label{eq:fit-PB}
\end{equation}
where $y_t=1/\nu$ is the thermal renormalization exponent, and $y_1$, $y_2$ are the leading and subleading correction exponents, respectively. 
The second-order term $q_2 (p_c-p)^2L^{2 y_t}$ is not present due to symmetry~\cite{noteVanish}. 
As a precaution against high-order correction terms that are not included in Eq.~(\ref{eq:fit-PB}), we gradually exclude the data points for $L\le L_{\rm min}$ and see how the residual $\chi^2$ changes with respect to $L_{\rm min}$. Generally the fit result is satisfactory if the value of $\chi^2$ is less than or close to the number of degrees of freedom (DF) and the drop of $\chi^2$ caused by increasing $L_{\rm min}$ is no more than one unit per degree of freedom.

For $z=8$, the fit results are summarized in Table~\ref{Tab:fit-PB}.
If letting all parameters of Eq.~(\ref{eq:fit-PB}) be free, the fitting procedure does not work,
which indicates that our MC data are not sufficient to determine all parameters simultaneously.
Therefore, we perform fits with some parameters being fixed.
When setting $b_2=0$, the fit results show that the leading correction exponent is $y_1 \simeq -3$.
In order to confirm this observation, we also perform the fits with $y_2$ being fixed at $-4$,
$-5$, or $-6$, but $b_2$ being free.
And the results are consistent with $y_1 \simeq -3$.
From these fits we also estimate $y_t=0.84(11)$ and $P_{{\rm B}0}=0.000\,007(8)$,
which are consistent with $y_t=3/4$~\cite{Nienhuis87} and $P_{{\rm B}0}=0$, as expected from universality of 2D ordinary percolation.
We further perform the fits with both $y_t=3/4$
and $P_{{\rm B}0}=0$ being fixed, which is helpful to give an accurate estimate of
$p_c$. 

\begin{figure}[htbp]
\centering
\includegraphics[scale=0.58]{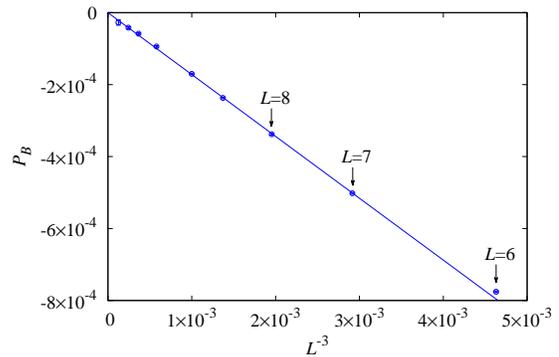}
	\caption{$P_B$ versus $L^{-3}$ for bond percolation on a periodic square lattice with $z=8$ equivalent neighbors, at $p=0.250\,368\,385$ which is within the error margin of the estimate $p_c=0.250\,368\,40(4)$. The solid line is a straight line with slope $b_1 \simeq -0.17$ obtained by fitting the data. From right (small) to left (large), sizes for other data points are $L=9,10,12,14,16,20$, respectively.}
\label{Fig:leadingcorrection}
\end{figure}

Thus, from all fits with $y_1$ free, we estimate 
the leading correction exponent of $\PB$ to be $y_1=-3.0(3)$.
And from all fits with $\PB$ fixed at zero, we report our estimate of
the percolation threshold as $p_c=0.250\,368\,40(4)$.
In Fig.~\ref{Fig:leadingcorrection}, we plot $P_B$ versus $L^{-3}$ for our MC data at $p=0.250\,368\,385$, which is within the error bar of our estimate of $p_c$. According to Eq.~(\ref{eq:fit-PB}), at $p_c$ and for large system sizes, $P_B$ versus $L^{y_1}$ should display approximately a straight line. This phenomenon is indeed observed in Fig.~\ref{Fig:leadingcorrection}, which demonstrates our estimate of $y_1 \simeq -3$. 
It is also noted that the magnitude of $P_{\rm B}$ is only of O$(10^{-5})$, illustrating the smallness of finite-size corrections in $P_{\rm B}$. 
We also perform fits for $R_2$ and $Q$ by adding $q_2 (p_c-p)^2 L^{2y_t}$ to Eq.~(\ref{eq:fit-PB}), 
which lead to estimates of the universal values $R_{2,0} = 0.309\,52(6)$ and $Q_0 = 0.960\,17(5)$, 
and the leading correction exponent $y_1 \simeq -1.6$.
The data of $R_2$ and $Q$ could also be fitted by formulae with more sophisticated 
finite-size corrections, e.g., with leading terms proportional to $L^{-2}$ and $\ln (L) L^{-2}$ 
for $R_2$, and for $Q$ with a term $\sim L^{-43/24}$ in addition to these two terms~\cite{Ouyang18}. 
These results of universal quantities are well consistent with the exact result 
$R_{2,0} = 0.309\,526\,28$~\cite{wrap2,wrap4} and with the previous 
estimate $Q_0 = 0.960\,17(1)$~\cite{ensemble}. From the estimate of the correction exponent $y_1$,
it is seen that the finite-size corrections for $P_{\rm B}$ decay more rapidly than those for $R_2$ and $Q$.

\begin{figure}
\begin{center}
\includegraphics[scale=1.0]{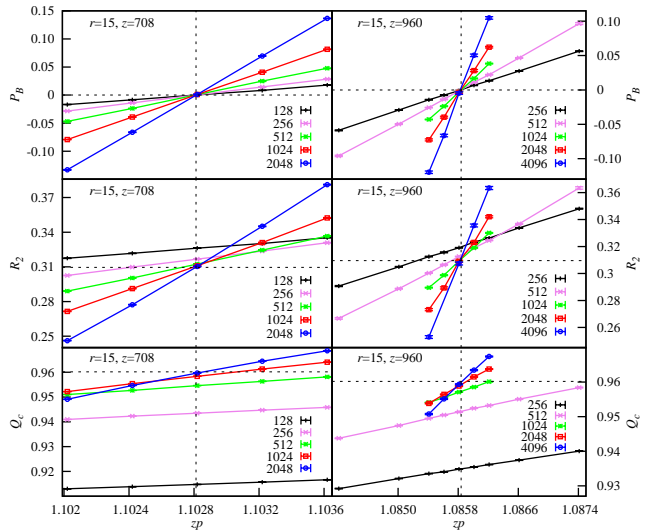}
\caption{
	Plots of $\PB$, $R_2$ and $Q$ versus $z p$ for different system sizes $L$, 
	for equivalent-neighbor percolation models with $r=15$, 
	corresponding to $z=708$ and $z=960$ for type-1 (left panel) 
	and type-2 (right panel) models, respectively.
	Vertical dashed lines show the thresholds
	$z p_c = 1.102\, 812(3)$ and $z p_c = 1.085\, 839(5)$ ~\cite{Deng19}
	for type-1 and type-2 models, respectively.
	Horizontal dashed lines indicate the universal values of these quantities at criticality.
	The error bars of the data are smaller than the size of the data points. 
	Values of $L$ are given in the legend. The solid lines connecting data points 
	are added for clarity.} 
\label{figure-r015}
\end{center}
\end{figure}

\begin{figure}
\begin{center}
\includegraphics[scale=1.0]{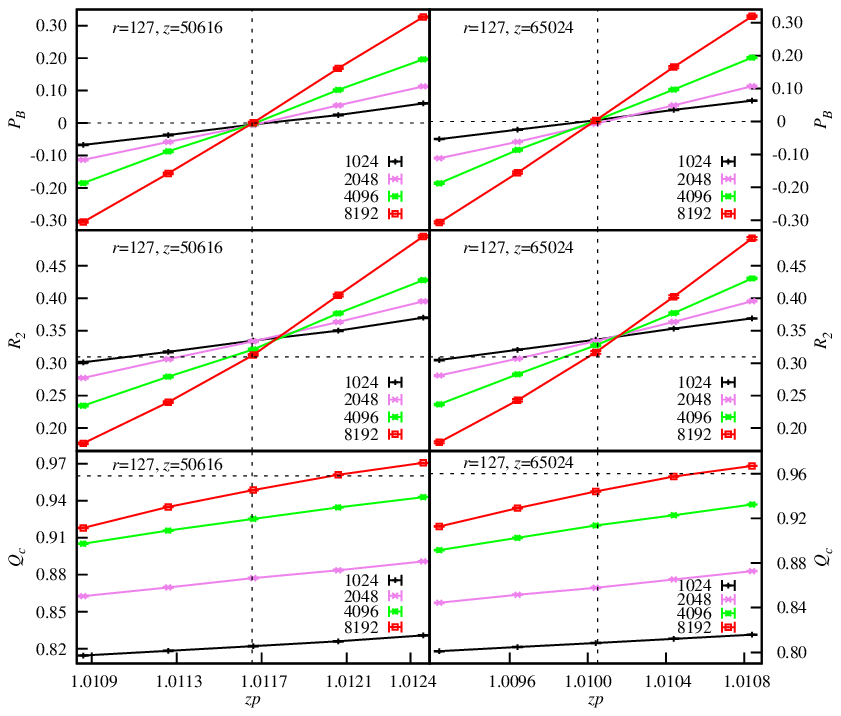}
\caption{
	Plots of $\PB$, $R_2$ and $Q$ versus $z p$ for different system sizes $L$, 
	for equivalent-neighbor percolation models with $r=127$, 
	corresponding to $z=50616$ and $z=65024$ for type-1 (left panel) 
	and type-2 (right panel) models, respectively.
	Vertical dashed lines show the thresholds
	$z p_c = 1.011\, 655(20)$ and $z p_c = 1.010\, 05(3)$ ~\cite{Deng19}
	for type-1 and type-2 models, respectively.
	Horizontal dashed lines indicate the universal values of these quantities at criticality.
	The error bars of the data are smaller than the size of the data points. 
	Values of $L$ are given in the legend. The solid lines connecting data points 
	are added for clarity.} 
\label{figure-r127}
\end{center}
\end{figure}

For models with $z>8$, we make plots for $\PB$, and compare them with 
those for other dimensionless quantities.
Figures~\ref{figure-r015} and \ref{figure-r127} show the results for $r=15$
and $127$, corresponding to $z \sim \text{O}(10^3)$ and O$(10^5)$, respectively.
We have the following observations.
Firstly, curves for different sizes $L$ cross well 
near the point $(z p_c\,, 0)$ for $\PB$, even for small relative sizes down to $L/(r+1)=8$.
Secondly, for $R_2$, as $L$ increases, the crossing points converge much slower than for $\PB$. 
For $r=127$, the convergence is so slow that even the crossing point of curves
for the largest two sizes deviates significantly from the critical point,
and if not knowing the exact value of $R_2$, a biased estimate of the critical point may be obtained.
Finally, for $Q$, the crossing point of the largest two sizes is significantly different 
from $(z p_c\,,Q_0)$ when $r=15$, and the curves do not intersect at all near $p_c$ when $r=127$.

Fits are also performed for models with $z > 8$ using Eq.~(\ref{eq:fit-PB}).    
For $\PB$, the leading correction exponent $y_1$ cannot be well determined 
when it is set as a parameter to be fitted.
With fixed $y_1 = -3$, stable fit results can be obtained, though the resulting 
estimate of $b_1$ has a large error bar that is comparable to its absolute value. 
These tell that our data are barely sufficient to detect the small finite-size correction in $\PB$.
The fit results also suggest that the second-order term $q_2(p_c - p)^2 L^{2 y_t}$ 
is absent in the scaling of $\PB$.
When fitting the data of $R_2$ and $Q$, the second-order term needs to be included.
For $R_2$ at $r=127$, if $R_{2,0}$ is not fixed in the fits,
the estimate of $p_c$ is significantly different from that obtained from fitting $\PB$, 
which confirms our second observation in last paragraph. 
For $Q$ at $r=127$, if $Q_0$ is not fixed, the estimate of $p_c$ is also biased, 
and the estimate of $Q_0$ is different from the universal value $0.960\,17(1)$; 
if $Q_0$ is fixed at the universal value, one cannot get stable fit results,
due to large and complicated finite-size corrections.
Thus $Q$ is not suitable for determining $p_c$ when $r$ (or equivalently $z$) is large, 
which is consistent with the previous observation for $Q$ that at $r=127$ curves for
different sizes do not intersect near $p_c$.

\begin{table}[htbp]
\begin{center}
\caption{Percolation threshold $z p_c$ for the equivalent-neighbor percolation
models of type-1 and type-2, with various number of neighbors $z$.
Results for $z>8$ have been reported in Ref.~\onlinecite{Deng19},
for which one of us (YD) is a coauthor.}
\label{Tab:zpc}
\begin{tabular}[t]{ l|ll|ll }
\hline
	\multicolumn{3}{c}{type-1} & \multicolumn{2}{c}{type-2} \\
\hline
	$r$ &	$z$	& $z p_c$ & $z$ & $z p_c$ \\  
\hline
	$1$	&	$4$ &  $2$ & $8$ & $2.002\,947\,2(32)$ \\
	$\sqrt{2}$	&	$8$ &  $2.002\,947\,2(32)$ & & \\
	$5$	&	$$ &  $$ & $120$ & $1.257\,695(7)$~\cite{Deng19} \\
	$7$	&	$148$ &  $1.234\,704(2)$~\cite{Deng19} & $224$ & $1.184\,443(5)$~\cite{Deng19} \\
	$15$	& 	$708$ & $1.102\,812(3)$~\cite{Deng19} & $960$ & $1.085\,839(5)$~\cite{Deng19} \\
	$23$	&	$1652$ & $1.066\,297(7)$~\cite{Deng19} & $2208$ & $1.055\,830(10)$~\cite{Deng19} \\
	$31$ 	&	$3000$ & $1.048\,803(8)$~\cite{Deng19} & $3968$ & $1.041\,349(7)$~\cite{Deng19} \\
	$35.8$	&	$4016$	& $1.042\,043(5)$~\cite{Deng19} & 	&	\\
	$47$	&	$6920$ & $1.031\,871(16)$~\cite{Deng19} & $9024$ & $1.027\,217(15)$~\cite{Deng19} \\
	$63$	& 	$12452$	& $1.023\,640(20)$~\cite{Deng19} & $16128$ & $1.020\,270(15)$~\cite{Deng19} \\
	$127$	&	$50616$	& $1.011\,655(20)$~\cite{Deng19} & $65024$ & $1.010\,05(3)$~\cite{Deng19} \\
\hline
\end{tabular}
\end{center}
\end{table}

\begin{figure}
\begin{center}
\includegraphics[scale=0.68]{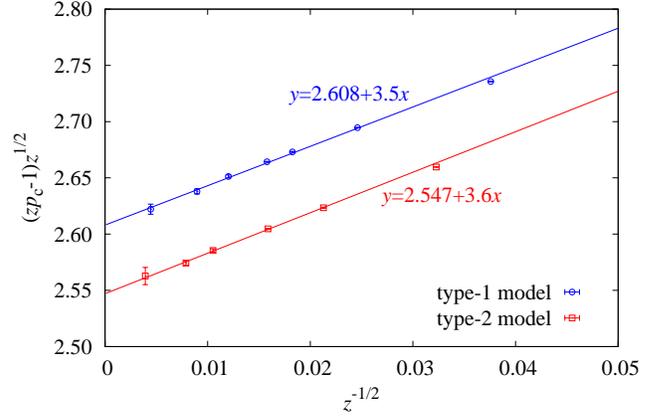}
\caption{
	Plot of $(z p_c - 1) z^{1/2}$ versus $z^{-1/2}$ for equivalent-neighbor models. The straight lines are obtained by fitting the data.}
\label{z-dependence}
\end{center}
\end{figure}

The above results demonstrate that $\PB$ also has much smaller 
finite-size corrections than other quantities when $z$ is greater than $8$.
And this advantage of $\PB$ becomes more obvious as $z$ increases. 
Thus we use $\PB$ to determine precisely percolation thresholds 
for various values of $z$ for both type-1 and type-2 models. 
The results are summarized in Table~\ref{Tab:zpc}. 
From the table, it can be seen that, when $z$ is large (e.g., $z>100$), 
the value of $z p_c$ decreases as $z$ becomes larger,
and it tends to approach the mean-field (MF) value $zp_c=z/(z-1)$
which equals to one in the limit $z \rightarrow \infty$.
Using these estimates of $z p_c$, we plot $(z p_c -1) z^{1/2}$ 
versus $z^{-1/2}$ for both types of models in Fig.~\ref{z-dependence}.
The intercept of the lines in the figure gives the value of $a_1$,
which is different for type-1 and type-2 models. 
The straight lines indicate that both models can be described 
by a correction term $a_2 z^{-1/2}$ when $z$ is large. 
Overall, the figure confirms that the threshold $p_c$ satisfies
 $z p_c -1 = a_1 z^{-1/2}(1 + a_2 z^{-1/2})$ when $z$ is large~\cite{Ouyang18,Deng19}.

For the asymptotic behavior of $z p_c$ as $z \rightarrow \infty$,
it has been conjectured that 
$z p_c - 1 \sim {1}/{r^{d-1}}$ for 2D and 3D models~\cite{Lalley14,Frei16},
where $d$ is the spatial dimension.
Since $z \sim r^d$, this leads to 
$z p_c - 1 \sim {1}/{z^{(d-1)/d}}$ for 2D and 3D models.
When $d=2$, it yields $z p_c - 1 \simeq a_1 z^{-1/2}$ for large $z$, 
which is supported by our results above.
Since $r^{d-1}$ is proportional to the surface length or area,
the asymptotic behavior of the form $a_1 z^{-1/2}$ can be regarded 
as a surface effect for the 2D model. 
Our observation that $a_1$  is different for the two types of models
also implies this surface effect, since the surfaces are different 
for type-1 and type-2 models. 

\section{Continuum percolation}
\label{sec-cont}

\subsection{Model and simulation}
\label{sec-simu}
Continuum percolation has been used to discuss the physical properties of complex fluids and disordered systems. The 2D continuum percolation with overlapping disks is particularly important because it corresponds to the randomly deposited networks of nanoparticles~\cite{nano}, which have various interesting properties and applications. In the 2D continuum percolation, a number ($n$) of randomly centered disks are distributed on a $L\times L$ square. The number $n$ satisfies a Poisson distribution 
\begin{equation}
P(n)=\frac{{\lambda^n} e^{-\lambda}}{n!} \,,
\end{equation}
where $P(n)$ refers to the probability that $n$ disks are distributed, and $\lambda = {\rho}L^2$ with $\rho$ being the mean density. Two penetrable disks are connected if they overlap, and the $n$ disks form connected groups with complex geometries. Various numerical studies have shown that continuum percolation with overlapping disks shares the same critical exponents with lattice percolation, indicating that they belong to the same universality class~\cite{uni1,uni2,uni3}.

We simulate the continuum percolation model
on a $L \times L$ square with periodic boundary conditions. 
The identical penetrable disks are of diameter one. 
In each trial, the number of objects $n$ is determined by a random number generator 
following a Poisson distribution with mean density parameter $\rho$. 
The disks are randomly placed into the square using a uniform distribution. 
The cell-list method~\cite{Frenkel2001} is employed for efficiently finding neighboring disks.
The same set of quantities as for the equivalent-neighbor model are sampled 
after all the clusters are constructed. 

As in site percolation, any pair of overlapping disks in continuum percolation can be considered to be effectively connected by a bond between their centers. One then obtains a nonplanar graph by drawing all such bonds between pairs of overlapping disks. However, for any pair of crossing bonds, the disks at their ends must belong to the same cluster. This is similar to site percolation on the square lattice with nearest- and next-nearest-neighboring interactions (coordination number $z=8$), for which four occupied sites on a square face, having a pair of diagonal bonds, must be in the same cluster. In other words, continuum percolation is like site percolation with compact neighborhoods where crossing connectivity cannot occur without simultaneously there being the presence of nearest-neighboring connectivity.
Actually the latter can be mapped to problems of lattice percolation of extended shapes (e.g., disks), whose thresholds can be related to the continuum thresholds for objects of those shapes~\cite{XunHaoZiff20}.
As a consequence, an interesting property arises for continuum percolation in 2D: the percolation of clusters and the void percolation of the unoccupied space are matching and if one percolates, the other does not, and vice versa. 

A recent numerical study of 
the continuum percolation of identical penetrable disks
was published by Mertens and Moore in 2012 ~\cite{MM}. 
In their work, wrapping probabilities are applied as observables, 
and an adaption of the Newman-Ziff algorithm is used for their simulations~\cite{wrap4,MM}.
They conduct extensive MC simulations for $50$ different system sizes ranging from $L=8$ to $2048$, 
with sample sizes being $10^{10}$ for $L \le 100$, $10^9$ for $100 < L \le 500$,
and $10^6$ for $500 < L \le 2048$.
In our work, we simulate $12$ different sizes ranging from $L=3$ to $512$. The number of samples is about $10^{10}$ for $L \le 100$ and $5 \times 10^9 $ for $100 \le L \le 512$. 
It is noted that, though not used in this work,
a similar Newman-Ziff approach as in Ref.~\onlinecite{MM}
can also be used to calculate $\PB$ as function of $\rho$, 
which might save some computer time since separate runs 
at different values of $\rho$ are not needed.

\subsection{Numerical results}

\begin{figure}
\begin{center}
\includegraphics[scale=1.6]{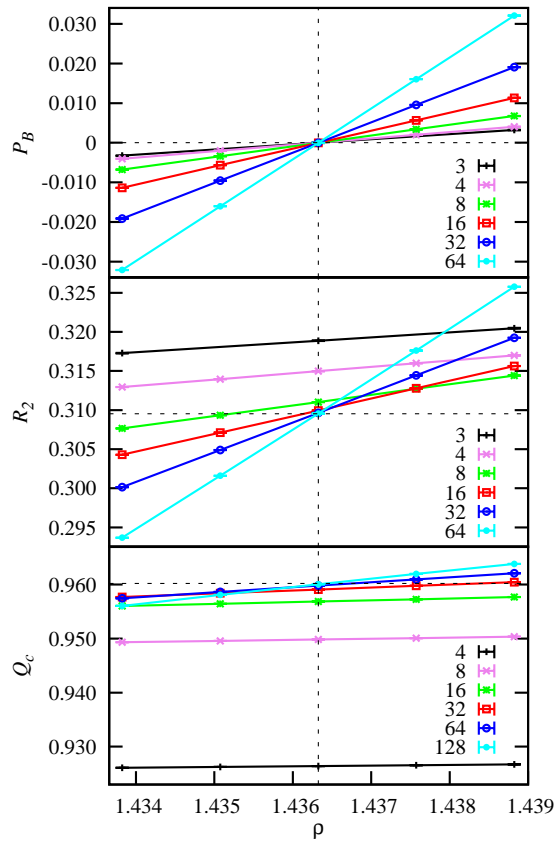}
\caption{
	Plots of $\PB$, $R_2$ and $Q$ versus $\rho$ for different system sizes $L$ 
	for the continuum percolation model. Values of $L$ are given in the legend. 
	Vertical dashed line shows the threshold $\rho_c = 1.436\, 325\, 05(10)$,
	and horizontal dashed lines indicate the universal values of these quantities at criticality.
	The error bars of the data are smaller than the size of the data points. 
	The solid lines connecting data points are added for clarity.} 
\label{figure-whole}
\end{center}
\end{figure}

Figure~\ref{figure-whole} shows the plots of quantities $\PB$, $R_2$ and $Q$ as a function of $\rho$ for different $L$. From the plot of $\PB$, it can be seen that the curves cross very well near $\rho \simeq 1.4363$, 
which is a rough approximation for the percolation threshold with an uncertainty at the fourth decimal place. At criticality, the value of $\PB$ is consistent with zero as expected from universality. 
For plots of $R_2$ and $Q$,
when $L$ is small, the curves cross at different points due to finite-size corrections.  
As $L$ becomes larger, the intersections of curves converge to the critical point, with $R_{2,0} \simeq 0.309$ and $Q_0 \simeq 0.960$ being consistent with 
their universal values $R_{2,0} = 0.309\,526\,28$~\cite{wrap2,wrap4} and $Q_0=0.960\,17(1)$~\cite{ensemble}. 

To examine the FSS behavior of sampled quantities, we fit the data by the ansatz
\begin{eqnarray}
	O(\rho,L) &=& O_0+a_1(\rho_c-\rho)L^{y_t}+a_2(\rho_c-\rho)^2L^{2 y_t} \nonumber \\
	& & +b_1L^{y_1}+b_2L^{2 y_1}+c_1(\rho_c-\rho)L^{y_t+y_1} \,,
\label{eq-fit}
\end{eqnarray}
where the thermal renormalization exponent is fixed at $y_t=3/4$.
For $\PB$ and wrapping probabilities,
the leading correction exponent $y_1$ is fixed as 
the subleading thermal renormalization exponent $-2$~\cite{Nienhuis87},
which is supported by previous data of wrapping probabilities for 2D continuum percolation~\cite{MM}. 
The fit results are shown in Tab.~\ref{fit}. 

\begin{table*}
\caption{Fit results of sampled quantities for the continuum percolation model.
	``Obs." is the abbreviation of ``observables".
	Entries ``--" indicate that the corresponding parameters are set to be zero, 
	and the numbers without error bars are fixed in the fits.} 
\label{fit}
\begin{center}
    \begin{tabular}{c|l|l|l|l|l|l|l|l|l|l}
    \hline
	    Obs. & $L_{\rm min}$ & ${\chi}^2$/DF & $O_0$ & $\rho_c$ & $a_1$ & $a_2$ & $b_1$ & $b_2$ & $c_1$ & $y_1$\\
 \hline
	    & $3$ & $95.3/102$ & $-0.000\,002(1)$ & $1.436\,324\,94(7)$ & $-0.567\,3(2)$ & $0.001(3)$ &  &  &  & \\
	    & $4$ & $95.3/99$ & $-0.000\,002(1)$ & $1.436\,324\,94(7)$ & $-0.567\,3(2)$ & $0.001(3)$ &  &  &  & \\
	    & $8$ & $88.2/94$ & $-0.000\,003(2)$ & $1.436\,324\,89(7)$ & $-0.567\,3(2)$ & $0.001(3)$ &  &  &  & \\
	    & $3$ & $100.5/103$ & $0$ & $1.436\,325\,05(5)$ & $-0.567\,3(2)$ & $0.001(2)$ &  &  &  & \\
      $\PB$ & $4$ & $100.4/100$ & $0$ & $1.436\,325\,05(5)$ & $-0.567\,3(2)$ & $0.001(2)$ & ~~~~-- & ~~~~-- & ~~~~-- & ~~~~--\\
	    & $8$ & $98.4/95$ & $0$ & $1.436\,325\,05(5)$ & $-0.567\,3(2)$ & $0.001(2)$ &  &  &  & \\
	    & $3$ & $100.7/104$ & $0$ & $1.436\,325\,05(5)$ & $-0.567\,3(2)$ & 0 &  &  &  & \\
	    & $4$ & $100.6/101$ & $0$ & $1.436\,325\,05(5)$ & $-0.567\,3(2)$ & 0 &  &  &  & \\
	    & $8$ & $98.6/96$ & $0$ & $1.436\,325\,05(5)$ & $-0.567\,3(2)$ & 0 &  &  &  & \\
    \hline
	    & $16$ & $72.0/69$ & $0.309\,526\,275$ & $1.436\,324\,88(7)$ & $-0.283\,6(2)$ & $0.052(2)$ & $0.118(2)$ & $-2.9(3)$ & $-0.02(4)$ & $-2$\\
	    $R_2$ & $24$ & $44.5/56$ & $0.309\,526\,275$ & $1.436\,324\,92(7)$ & $-0.283\,4(2)$ & $0.052(2)$ & $0.123(3)$ & $-7(2)$ & $-0.2(3)$ & $-2$\\
	    & $32$ & $37.3/51$ & $0.309\,526\,275$ & $1.436\,324\,97(8)$ & $-0.283\,5(2)$ & $0.052(2)$ & $0.131(7)$  & $-16(7)$ & $-0.3(3)$ & $-2$\\

    \hline
	    & $16$ & $65.2/69$ & $0.309\,526\,275$ & $1.436\,324\,95(7)$ & $\,\,\,\,0.283\,5(1)$ & $0.056(2)$ & $0.125(1)$  & $-4.5(3)$ & $\,\,\,\,0.04(4)$ & $-2$\\
	    $R_0$ & $24$ & $37.5/56$ & $0.309\,526\,275$ & $1.436\,324\,92(7)$ & $\,\,\,\,0.283\,2(2)$ & $0.056(2)$ & $0.129(3)$ & $-8(2)$ & $\,\,\,\,0.6(2)$ & $-2$\\
	    & $32$ & $36.3/51$ & $0.309\,526\,275$ & $1.436\,324\,91(8)$ & $\,\,\,\,0.283\,2(2)$ & $0.056(2)$ & $0.131(7)$ & $-9(7)$ & $\,\,\,\,0.6(2)$ & $-2$\\

    \hline
	    & $64$ & $34.2/33$ & $0.960\,173(4)$ & $1.436\,327(1)$ & $-0.040\,89(3)$ & $0.014\,5(1)$ & $-0.20(2)$ &  & $-0.04(2)$ & $-1.51(2)$ \\
	    $Q$ & $96$ & $19.3/24$ & $0.960\,176(7)$ & $1.436\,327(2)$ & $-0.040\,87(4)$ & $0.014\,5(2)$ & $-0.18(4)$ & ~~~~-- & $-0.08(7)$ & $-1.48(5)$ \\
	    & $128$ & $14.4/20$ & $0.960\,2(1)$ & $1.436\,337(8)$ & $-0.040\,83(6)$ & $0.014\,5(2)$ & $-0.03(3)$ &  & $-0.01(1)$ & $-1.0(3)$ \\
    \hline
    \end{tabular}
\end{center}
\end{table*}

\begin{figure}
\begin{center}
\includegraphics[scale=0.68]{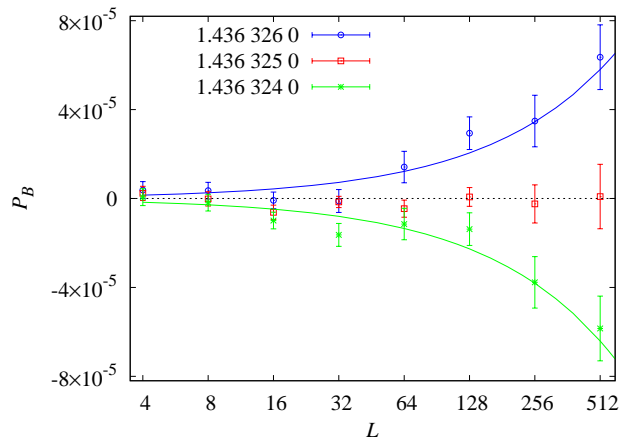}
\caption{
Plot of $\PB$ versus $L$ at different mean densities $\rho$ near criticality. Standard re-weighting technique is applied to obtain the data.  Values of $\rho$ are given in the legend.  The curves are obtained by fitting the data.  } 
\label{figure-narrow}
\end{center}
\end{figure}

For $\PB$, the amplitudes $b_1$, $b_2$ and $c_1$ are found to be consistent with zero when they are set as parameters to be fitted, which indicates that the finite-size correction is very small. The presented results for $\PB$ are from fits with $b_1$, $b_2$ and $c_1$ being fixed at zero. 
When $O_0$ is a free fit parameter, the fitted values of $O_0$ for $\PB$ is consistent with zero within one error bar, as expected from the universality of $\PB$. Then fits are performed with fixed $O_0=0$.  
It is found that, with only the second and third terms, Eq.~(\ref{eq-fit}) can well describe the $\PB$ data for $L\ge 3$ near the critical point, yielding a stable estimate of $\rho_c$ as $1.436\,325\,05(5)$. Moreover, the fit results have $a_2$ being consistent with zero, which implies that the second-order term $a_2 (\rho_c-\rho)^2L^{3/2}$ vanishes also due to symmetry~\cite{noteVanish}. 
Fits with fixed $a_2=0$ also lead to the estimate of $\rho_c$ as $1.436\,325\,05(5)$.
Thus we set our final estimate as $\rho_c = 1.436\,325\,05(10)$, where the error bar 
is quoted as twice the statistical error to account for possible systematic errors.
The systematic errors may be due to higher-order scaling terms or the very small finite-size correction not included in the fits. 
Figure~\ref{figure-narrow} shows a plot of $\PB$ versus $L$ at three different values of $\rho$ that are very close to the critical point. It is found that the data points at $\rho_c \simeq 1.436\,325\,0$ distribute around $\PB=0$ regardless of the system size $L$, i.e., the finite-size correction in $\PB$ is undetectable at criticality. 
The obvious deviation from $\PB=0$ when $\rho \ne \rho_c$  illustrates the reliability of our estimate of $\rho_c$.

For  $R_2$ and $R_0$, the value of $O_0$ is fixed at the theoretical predictions in the fitting. The data up to $L_{\rm min}=16$ have to be discarded for a reasonable residual $\chi^2$. The results support the presence of the leading correction term $\sim L^{-2}$ with the amplitude $b_1 \simeq 0.12$. Together with the fact that the coefficient $a_1$ of $R_2$ and $R_0$ have the same amplitude but opposite signs, it is suggested that $R_2(\epsilon) = R_0(-\epsilon)$ with $\epsilon = (\rho_c - \rho)L^{y_t}$, 
which is expected from duality~\onlinecite{noteVanish}.

For $Q$, as seen from Fig.~\ref{figure-whole},
the finite-size correction is much larger than that in 
$R_2$ and $\PB$. 
When the data are fitted to Eq.~(\ref{eq-fit}), the coefficient $b_2$
has an error bar much larger than the central value. 
Thus fits are performed with fixed $b_2=0$.
A large cut-off $L_{\rm min}=64$ has to be set for a stable fit. 
The results show a leading correction term with exponent $y_1 \simeq -1.5$.

\section{Discussion and conclusion}
\label{sec-dis}

In summary, we study the critical polynomial $\PB$ in nonplanar and continuum percolation models
by MC simulations and FSS analysis. Two kinds of models are considered, 
i.e., the bond percolation model on square lattice with many equivalent neighbors (a nonplanar model) 
and the 2D continuum percolation of identical penetrable disks.
Similar to properties observed in planar-lattice models, 
it is found for these two models that $\PB=0$ holds at the critical point
as expected from universality, and that the finite-size correction in $\PB$ 
is very small.

For $\PB$ in the 2D equivalent-neighbor percolation model, from the data of 
the model with $z=8$ neighbors, we find that the leading correction exponent is $y_1 \simeq -3$,
smaller than those for the wrapping probability and the dimensionless ratio
related to the cluster-size distribution.
The advantage of $\PB$ over other quantities is more significant as $z$ increases.
Thus, for two types of equivalent-neighbor models with different ways to involve neighbors, 
$\PB$ is employed to determine precisely the percolation threshold $p_c(z)$ 
for various values of $z$. 
The asymptotic behavior of $z p_c$ is confirmed to be $z p_c -1 \simeq a_1 z^{-1/2}$ 
for $z \rightarrow \infty$, with the coefficient $a_1$ being different for 
the two types of models. Since the regions of neighbors have different surfaces 
for the two types of models, the observed difference of $a_1$ could be regarded 
as evidence that the term $a_1 z^{-1/2}$ is a surface effect~\cite{Lalley14,Frei16}. 
We also find that the subleading dependence of $z p_c -1$ on $z$
is proportional to $z^{-1}$.

Equivalent-neighbor percolation models have also been studied in more than two dimensions in the literature.
For $d=3$,  while the implied surface effect suggests the $z \rightarrow \infty$ asymptotic behavior 
$z p_c - 1 \simeq {a_1}{ z^{-2/3}}$~\cite{Lalley14, Frei16},
a most recent numerical study finds empirically 
$z p_c - 1 \simeq {a_1}{ z^{-1/2}}$~\cite{Xun20}.
Since the maximum value of $z$ considered in Ref.~\onlinecite{Xun20} is
$146$, it would be interesting to simulate systems with much larger $z$
to clarify the ambiguity of the correction exponent. 
For $d \ge 4$, it is suggested that
$z p_c - 1 \simeq {a_1}/{z}$ (with logarithm corrections in $d=4$)~\cite{Frei16, Hofstad05},
which implies that in this case the asymptotic behavior of 
$z p_c$ is a bulk property.
More work is needed to confirm the above asymptotic behavior for $d \ge 4$,
and to understand the difference of the correction exponents in different dimensions.

For $\PB$ in the 2D continuum percolation model, 
it is found that the finite-size correction is undetectable for $L \ge 3$. 
Thus by using $\PB$, we are able to determine precisely 
the continuum percolation threshold as $\rho_c = 1.436\, 325\, 05(10)$. 
This estimate is slightly below the previous value $\rho_c = 1.436\, 325\, 45(8)$ 
obtained by analyzing the FSS of wrapping probabilities~\cite{MM}.
Our simulations are with smaller system sizes 
than the previous work as described in Sec.~\ref{sec-simu}, but the resulting 
error bars of $\rho_c$ are of the same order, i.e., $10^{-7}$. 

\begin{figure}
\begin{center}
\includegraphics[scale=0.68]{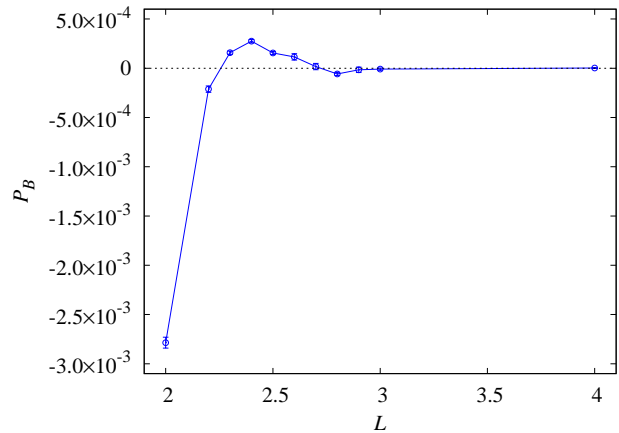}
\caption{
	Plot of $\PB$ versus $L$ for $L \le 4$ in continuum percolation, 
	at $\rho = 1.436\,325\,0$ that is within the error margin of 
	the estimated critical point $\rho_c = 1.436\,325\,05(10)$.
	The line connecting data points is added for clarity.} 
\label{fractional}
\end{center}
\end{figure}

As mentioned in the introduction, for unsolved planar-lattice percolation models at criticality, $\PB$ usually has a leading correction term that scales as $\sim L^{-3.25}$($\Delta=4$) or $\sim L^{-5.25}$($\Delta=6$)~\cite{j3}; and for exactly solvable lattice percolation problems, the finite-size correction in $\PB$ vanishes for arbitrary size $L$.
Might the continuum model be similar to the exactly solvable lattice models also for system sizes $L<3$ ?
To answer this,
since $L$ is not limited to integers, we perform additional simulations for system sizes 
$2 \le L < 3$ at $\rho_c$.
The result is shown in Fig.~\ref{fractional}. A nonzero correction is observed for $L \le 2.8$, which means that the finite-size correction in $\PB$ does not vanish for arbitrary $L$, although it is negligible for $L \ge 3$. 
It is nevertheless surprising to see that the amplitude of finite-size corrections is small and in order O($10^{-3}$) even for $L=2$.

Why the finite-size correction in $\PB$ is so small in the 2D continuum percolation model remains an open question. 
For exactly solved lattice percolation models, the symmetry of the lattice can lead to the absence of correction terms in the FSS of $\PB$, which is proved by Mertens and Ziff~\cite{MZ} on self-dual lattices and self-matching lattices. 
Our results support that, in the continuum percolation model for $L \ge 3$, $\PB$ is antisymmetric around $p_c$, 
which exactly holds for bond percolation on self-dual lattices~\cite{noteVanish}. 

The critical polynomial $\PB$ can also be applied to study 
the continuum percolation of other shaped objects, 
the nonplanar Potts model in the FK representation etc.
$\PB$ is currently defined in two dimensions.
In more than two dimensions, one can also define various types
of wrapping probabilities according to their topological properties.
Is it possible to define a quantity similar to $\PB$ from the combination of 
these wrapping probabilities? 
With the great success of the application of $\PB$ in two dimensions,
it is very attractive to explore the possibility. 
If found, the quantity could have many applications, such as helping clarify 
the $z$-dependence of $z p_c -1$ for equivalent-neighbor percolation models 
with $d \ge 3$.

\acknowledgments
We thank R. Ziff for very helpful comments.
H. H. acknowledges the support by the National Science Foundation of China (NSFC) under Grant No.~11905001,
and by the Anhui Provincial Natural Science Foundation of China under Grant No.~1908085QA23.
J. F. W. acknowledges the support by the NSFC under Grant No.~11405039.
Y. D. acknowledges the support by the National Key R\&D Program of China under Grant No.~2016YFA0301604 
and by the NSFC under Grant No.~11625522.


\end{document}